\documentclass[final,1p,twocolumn]{elsarticle}

\usepackage{graphicx}
\usepackage{amssymb}

\biboptions{sort&compress} 

\usepackage[hyphens]{url}

\usepackage{xcolor}

\usepackage{soul}
\usepackage{comment}
\usepackage{lineno}

\journal{Nuclear Instruments and Methods in Physics Research Section A}

\begin{document}

\begin{frontmatter}


\title{The High-Altitude Water Cherenkov (HAWC) Observatory in M\'exico: The Primary Detector \footnote{Article accepted. Please cite the article as:  A.U. Abeysekara,A. Albert,R. Alfaro et al., The High-Altitude Water Cherenkov (HAWC) observatory in México: The primary detector, Nuclear Inst. and Methods in Physics Research, A (2023) 168253, https://doi.org/10.1016/j.nima.2023.168253}}




 

\author[UU]{A.U.~Abeysekara}

\author[LANL]{A.~Albert}

\author[IF-UNAM]{R.~Alfaro}

\author[UNACH]{C.~Alvarez}

\author[UMSNH]{J.D.~Álvarez}

\author[UCR]{M.~Araya}

\author[UMSNH]{J.C.~Arteaga-Velázquez}

\author[IGeof-UNAM,Kerala]{K.P.~Arunbabu}

\author[IF-UNAM]{D.~Avila Rojas}

\author[PSU]{H.A.~Ayala Solares}

% ...

\address[UU]{Department of Physics and Astronomy, University of Utah, Salt Lake City, UT, USA }

\address[LANL]{Physics Division, Los Alamos National Laboratory, Los Alamos, NM, USA }

\address[IF-UNAM]{Instituto de F\'{i}sica, Universidad Nacional Autónoma de M\'exico, Ciudad de M\'exico, M\'exico }

\address[UNACH]{Universidad Autónoma de Chiapas, Tuxtla Gutiérrez, Chiapas, M\'exico}

\address[UMSNH]{Universidad Michoacana de San Nicolás de Hidalgo, Morelia, M\'exico }

\address[UCR]{Escuela de F\'{i}sica, Universidad de Costa Rica, San Jos\'e, Costa Rica }

\address[IGeof-UNAM]{Instituto de Geof\'{i}sica, Universidad Nacional Autónoma de M\'exico, Ciudad de M\'exico, M\'exico }

\address[Kerala]{Department of Physics, St. Albert$'$s College (Autonomous), Cochin, 682018 Kerala, India}

\address[PSU]{Department of Physics, Pennsylvania State University, University Park, PA, USA }

% ...
\address[FC-UNAM]{Facultad de Ciencias, Universidad Nacional Autónoma de M\'exico, 04510, Ciudad de M\'exico, M\'exico}

% ...
\address[IA-UNAM]{Instituto de Astronom\'{i}a, Universidad Nacional Autónoma de M\'exico, Ciudad de M\'exico, M\'exico }

% ...
\address[ICN-UNAM]{Instituto de Ciencias Nucleares, Universidad Nacional Autónoma de M\'exico, Ciudad de M\'exico, M\'exico }

% ...
\address{{\bf The historical and present HAWC Collaboration}\footnote{Principal corresponding author: Eduardo de la Fuente (eduardo.delafuentea@academicos.udg.mx)},\footnote{Corresponding authors: Jim Linnemann, Wayne Springer, Ibrahim Torres, \& Hao Zhou} \\

 }

\begin{abstract}
\label{abstract}

The High-Altitude Water Cherenkov (HAWC) observatory is a second-generation continuously operated, wide field-of-view, TeV gamma-ray observatory. The HAWC observatory and its analysis techniques build on experience of the Milagro experiment in using ground-based water Cherenkov detectors for gamma-ray astronomy.
HAWC is located on the Sierra Negra volcano in M\'exico at an elevation of 4100 meters above sea level. The completed HAWC observatory principal detector (HAWC) consists of 300 closely spaced water Cherenkov detectors, each equipped with four photomultiplier tubes to provide timing and charge information to reconstruct the extensive air shower energy and arrival direction. The HAWC observatory has been optimized to observe transient and steady emission from sources of gamma rays within an energy range from several hundred GeV to several hundred TeV. However, most of the air showers detected are initiated by cosmic rays, allowing studies of cosmic rays also to be performed. This paper describes the characteristics of the HAWC main array and its hardware.

\end{abstract}



\end{frontmatter}






\section{Introduction}
\label{sec:intro}

Very-high-energy (VHE) gamma rays (E$>$100 GeV) probe the non-thermal universe, tracing the sites of particle acceleration around black holes, neutron stars, astrophysical jets, massive star formation regions, and other objects where strong shocks are present. While human-made particle accelerators can currently accelerate particles to a few tens of TeV, nature can accelerate particles to at least 10$^{8}$ TeV (e.g. \cite{DeAngelis2018, Sinnis2021}). Gamma rays have been observed to energies above 100 TeV (e.g. \cite{LhPevb}). VHE gamma rays can be produced by inverse Compton scattering, synchrotron and bremsstrahlung radiation, and the decay of neutral pions created by collisions of accelerated hadrons with an ambient medium or radiation fields (e.g. \cite{DeAngelis2018,Aharonian2004} and references therein).  VHE gamma-ray observations shed light on environments where charged particles are accelerated, thereby gaining insight into some of the most extreme regions of our universe. From radio to VHE gamma rays, multi-wavelength observations create a detailed picture of these acceleration environments. More recently, multi-messenger observations of gamma rays, neutrinos, and gravitational radiation have been used to improve our understanding of mergers of black holes and neutron stars~\cite{Abbott2017}.

\begin{figure*}[htb!]
\centering\includegraphics[width=\textwidth]{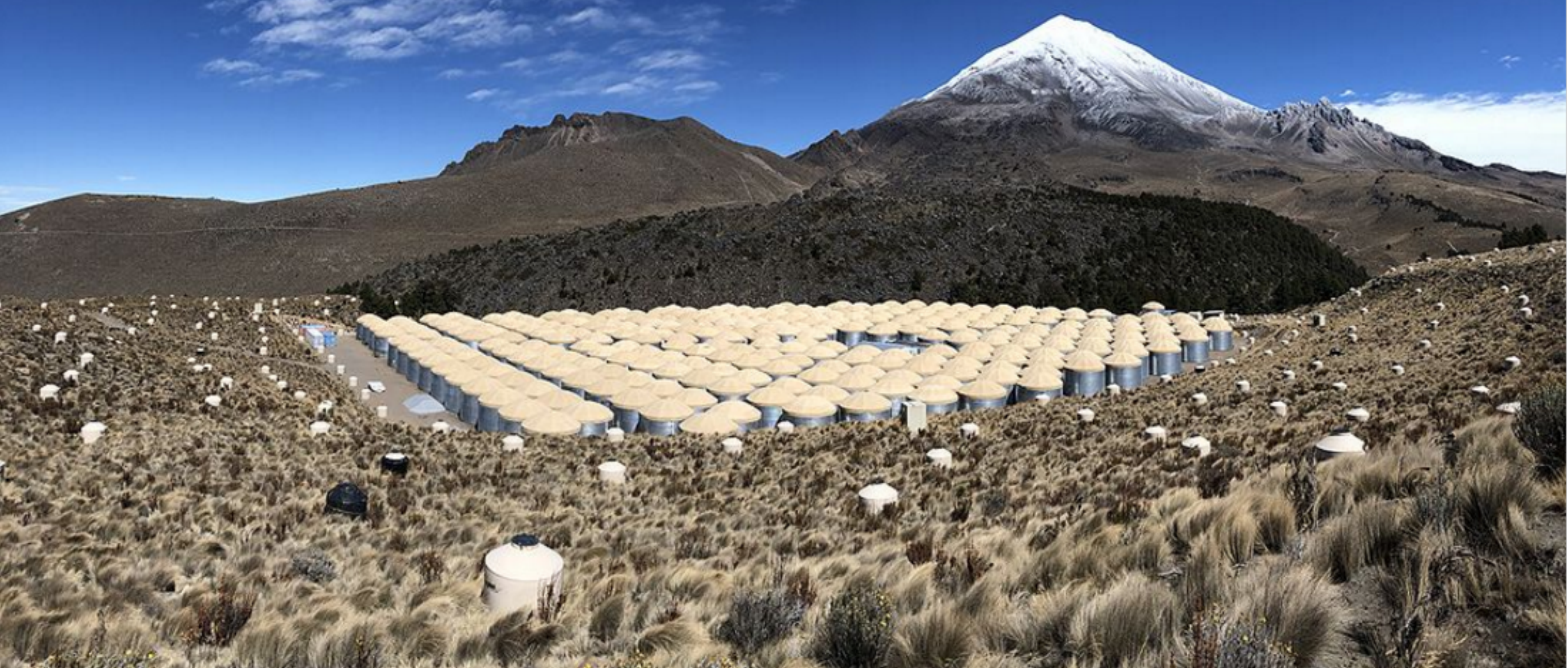}
\caption{The HAWC observatory, seen looking north from the Sierra Negra volcano. The Citlaltepetl volcano is visible in the background. The primary detector (HAWC) is at the center, surrounded by the smaller plastic tanks (outriggers) not discussed in this paper (see Section \ref{S:Future}). Credit: The HAWC Collaboration.} 
\label{fig:hawc_obs}
\end{figure*}

The Earth's atmosphere is opaque to X-rays and gamma rays. At sufficiently high energies, gamma and cosmic rays can generate extensive air showers (EAS), with a pancake-shaped wavefront of relativistic particles (mostly electrons, positrons, and lower-energy photons from primary gamma rays, and pions and muons from cosmic rays). The EAS generates Cherenkov light in the atmosphere (detected by atmospheric Cherenkov telescopes; e.g. \cite{Aharonian2008, DiSciascio2019} and references therein), and if the initial gamma ray has sufficiently high energy, particles in the EAS will survive to ground level and can be detected by an array of detectors on the ground. Thus, there are three (overlapping) gamma-ray energy regimes depending on the detection techniques. At low energies, space-based instruments (e.g. Fermi--LAT; \cite{FermiLAT2009}) are used to detect the primary gamma rays from $\sim$50 MeV to $\sim$100 GeV, atmospheric Cherenkov telescopes work from $\sim$50 GeV to $\sim$10 TeV and particle detection arrays operate from $\sim$1 - $\sim$1000 TeV \cite{LhPeva,LhPevb}. In this paper, we discuss the High Altitude Water Cherenkov (HAWC) TeV gamma-ray observatory, which uses water Cherenkov detectors (WCDs) to detect the particles within an EAS that survive to ground level at high altitude. The HAWC observatory was designed to extend the capabilities of the first-generation water Cherenkov TeV gamma-ray  observatory, Milagro (\cite{Atkins2004, Abdo2007, Abdo2014} and references therein). HAWC became the first EAS gamma-ray detector to observe dozens of sources at energies from one TeV to above 100 TeV (e.g \cite{Albert2020g} and references therein). For a review of WCDs for gamma-ray astronomy, EAS, and Cherenkov radiation, see \cite{Sinnis2021} (and references therein).

The HAWC observatory is located in the state of Puebla, M\'exico in the  Pico de Orizaba National Park at an altitude of 4100 m.a.s.l. (meters above sea level)\textcolor{red}{,} in the saddle region between the Sierra Negra volcano, or Tliltepetl (altitude 4582 m.a.s.l.), and Pico de Orizaba, or Citlaltepetl (``star mountain'' in Nahuatl; altitude 5636 m.a.s.l. Figure~\ref{fig:hawc_obs} shows a picture of the location of the HAWC observatory with  Citlaltepetl in the background. The Gran Telescopio Milim\'etrico (GTM) Alfonso Serrano or Large Millimeter Telescope\footnote{\url{http://lmtgtm.org/telescope/telescope-description/}} (LMT; \cite{Hughes2020}) site is at the summit of Sierra Negra. The facilities and infrastructure of the nearby LMT have been leveraged for the benefit of the HAWC observatory. The HAWC observatory is comprised of the original primary detector (HAWC hereafter) and an upgrade consisting of 345 outrigger detectors\footnote{This paper does not focus on the outriggers, only on HAWC.} (see Section \ref{S:Future}; \cite{Marandon2021} and references therein). The facilities also include a central Counting House (CH) with  electronics  and computers, and a support building, called the HAWC Utility Building (HUB), containing water purification facilities and an office. More details about the development of the site, history, and origin of the HAWC observatory are presented in \cite{Carraminana2008,Torres2011,Gonzalez2011,Gomez2019}.

The latitude of the HAWC observatory is 18.99$^{\circ}$ North. Taking  50$^{\circ}$ from zenith as the limit of the viewable field,  we are able to 
observe gamma-ray sources to declination to at least 31$^{\circ}$ South (including the Galactic center).  In a sidereal day, we observe up to  9 sr ($>70\%$ of the entire sky). Observations at larger zenith angles are possible, but the sensitivity declines, and the energy threshold increases rapidly with further increasing inclination due to increased atmospheric depth and shorter transits. The 97.31$^{\circ}$ W longitude of the HAWC observatory is similar to that of major observatories in M\'exico, the United States, and South America. This location facilitates prompt multi-wavelength follow-up observations of transients found in HAWC data. 

The observatory was constructed with financial support led by the Consejo Nacional de Ciencia y Tecnolog\'ia (CONACyT) of M\'exico, and the United States National Science Foundation (NSF) and Department of Energy (DoE). 

This paper focuses on the hardware of the HAWC main array. A forthcoming paper will describe the outrigger array (additional small tanks to improve HAWC performance at high energy). In Section~\ref{S:HMDandOp}, we present an overview of HAWC, including a history of its construction and a summary of HAWC science topics. In Section~\ref{S:site}, we present a description of the instrument and its construction. The electronics, calibration, and operations are discussed in Sections \ref{S:Electronics}, \ref{S:Calibration}, and \ref{S:Operations}, respectively. The future of the HAWC observatory and its recent outrigger upgrade is briefly discussed in Section~\ref{S:Future}.

\section{The HAWC Primary Detector: Overview,  Installation History, and Science}
\label{S:HMDandOp}

\begin{figure}[ht]
\centering\includegraphics[width=0.5\textwidth]{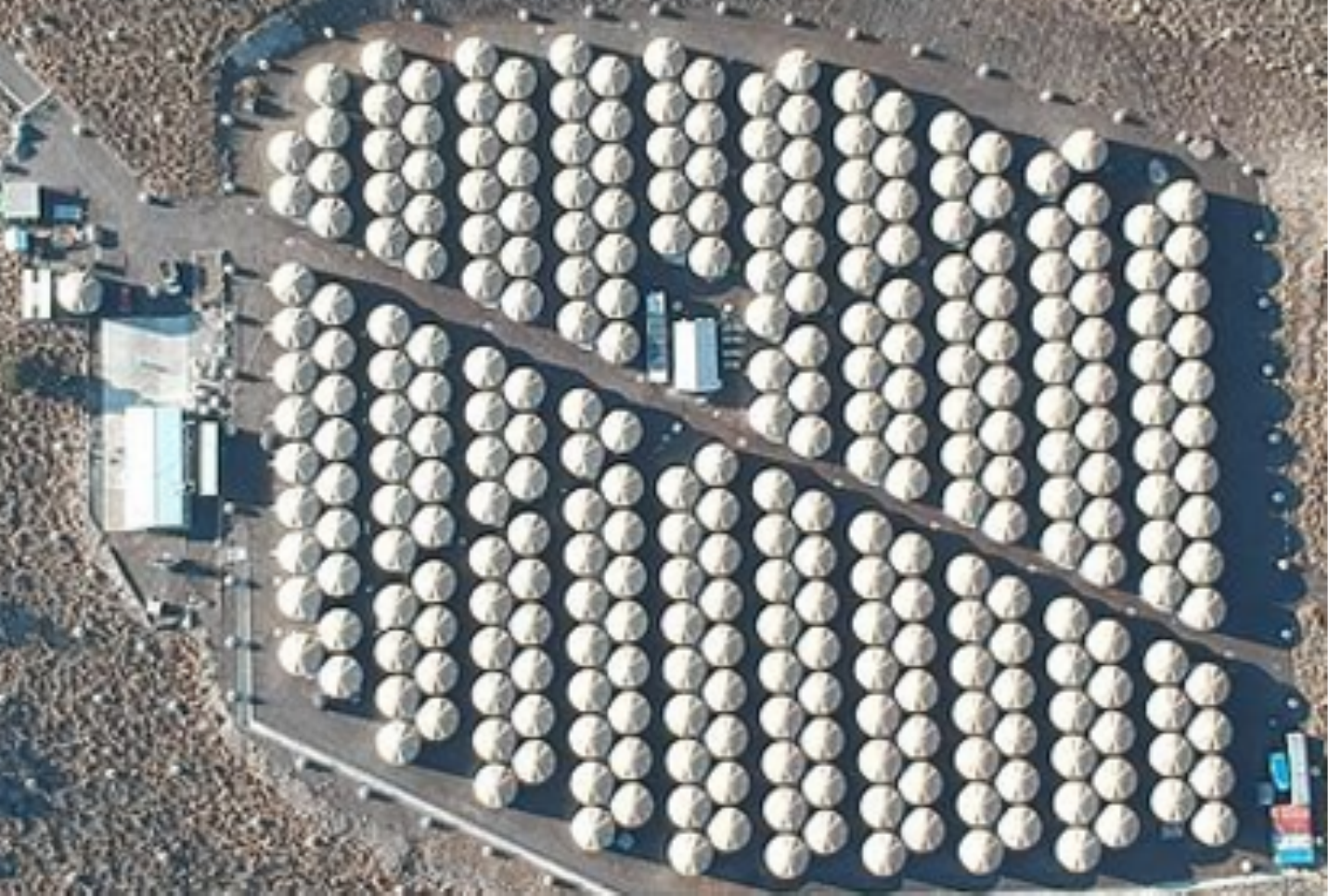}
\centering\includegraphics[width=0.5\textwidth]{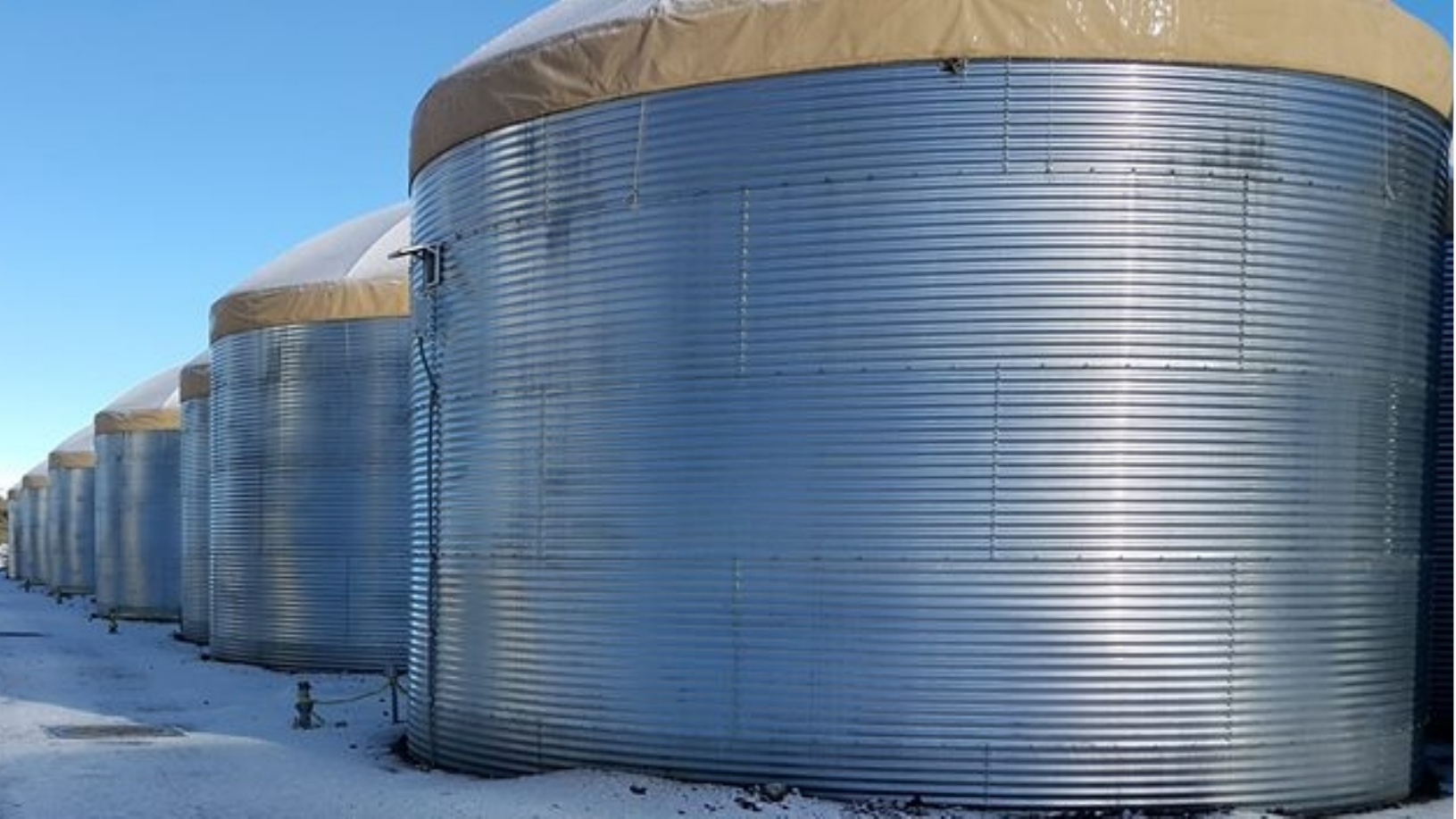}
\centering\includegraphics[width=0.5\textwidth]{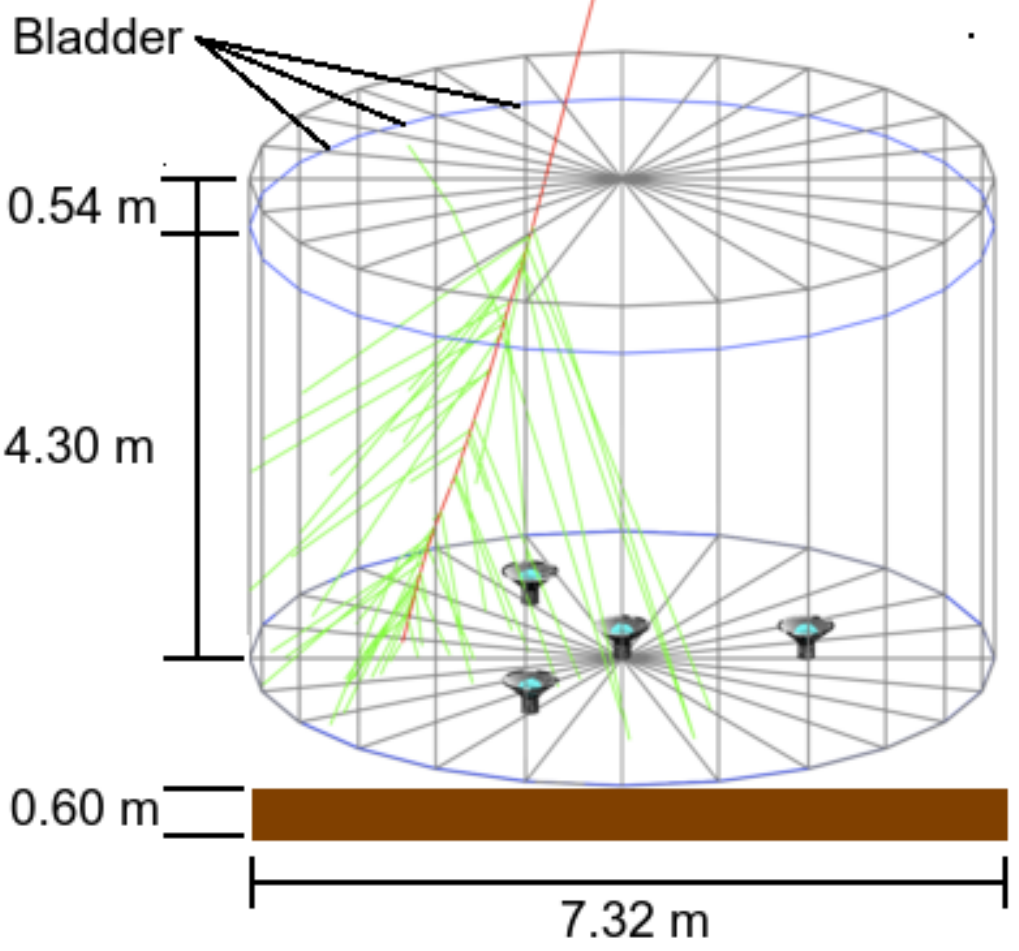}
\caption{\textbf{Top}: Drone view of the HAWC primary detector, a dense array of 300 WCDs. The counting house (CH) is in the top-center. The HAWC Utility Building (HUB) is on the left. \textbf{Center}: View down the diagonal truck access road through the HAWC WCDs. The segmented panels of the tank can be seen. \textbf{Down}: Schematic of a WCD. The red line represents an incoming muon, and the green lines represent thinned Cherenkov light. Three eight-inch PMTs are located around a central ten-inch PMT. See text for details.}

\label{fig:HMD_aerial}
\end{figure}

HAWC is an array of 300 steel water tanks equipped with large-area photomultiplier tubes (PMTs). These WCDs cover an area of over 22,000 m$^2$ (Figure~\ref{fig:HMD_aerial} top), five times that of the top layer of Milagro and about ten times larger than its bottom layer\footnote{The bottom layer of Milagro was used to reject the hadronic cosmic-ray background.}. The combination of increased array area, higher altitude (4100 m.a.s.l. vs. 2630 m.a.s.l.), optical isolation of the HAWC WCDs, and all HAWC PMTs being located at an intermediate depth of 4 m, results in a sensitivity increase of more than one order of magnitude when compared to Milagro \citep{Abdo2007,Abeysekara2019a}. HAWC's vertical atmospheric depth is 637 g cm$^{-2}$, equivalent to 17.4 radiation lengths. The reduced overburden compared to Milagro (20.8 radiation lengths) allows HAWC to sample air showers more than 3 radiation lengths earlier (nearer to shower maximum) allowing more particles to arrive at the detector. This lowers HAWC's energy threshold and improves reconstruction accuracy and hadron rejection.   

We flattened the HAWC platform, with a slope of approximately 1.0\%, enough for the drainage of precipitation.  The downward direction of the slope points 15.8\textdegree ~West of North along the direction of the tank columns. Consequently, the normal to the platform is tilted from the zenith along the direction of this slope by approximately 0.7\textdegree. 

The WCDs (see Figure~\ref{fig:HMD_aerial} center) are cylindrical steel structures ($\sim$ 7.3 m in diameter and $\sim$ 5.4 m in height that support a light-tight bladder containing $\sim$ 180,000 liters of purified water (see Section \ref{bladder}). The water depth is 4.5 m. The minimum separation between two adjacent WCDs is 60 cm, sufficient to service the entire exterior of a tank. There is an aisle 2.6 m wide between each column pair to allow passage of a standard-width truck.  A 4.3m access road, known as 10th Avenue, runs diagonally across the array. Cable trenches with conduits and pull-boxes were placed in 10th Avenue  as well as the narrower diagonal paths visible in Figure ~\ref{fig:HMD_aerial}.

HAWC reuses 900 Hamamatsu R5912 eight-inch PMTs \cite{hamamatsu} and 1200 front-end electronic channels from Milagro, with a newly developed Data Acquisition (DAQ) System. An additional 300 larger Hamamatsu R7081 ten-inch PMTs \cite{hamamatsu} with higher quantum efficiency (30\%) were added to increase sensitivity to lower-energy gamma rays (see Section \ref{sec:PMTs}). Thus, every WCD has four upward facing PMTs anchored by a mounting system. The ten-inch PMT at the center is surrounded by three eight-inch PMTs in an equilateral triangle configuration. The PMTs are $\sim$ 0.5 m above bottom (one radiation length = 0.4 m). The distance from the eight-inch PMTs to the central ten-inch PMT is 1.83 m, midway between the center and the outer wall (see Figure~\ref{fig:HMD_aerial} down).

The $\sim$ 200 m $\times$ 150 m HAWC layout shown in the top panel of Figure~\ref{fig:HMD_aerial} was simulated using HAWCSim. HAWCSim contains a detailed description of the HAWC array and propagates the air shower particles provided by CORSIKA \cite{Heck1998} at the top of the WCDs.  It uses GEANT 4 \cite{Agostinelli2003}, including the generation of Cherenkov photons and the response of the PMT's down to a single photoelectron. A detailed description of the HAWC design can be found in \cite{Smith2009}. We chose HAWC's  final layout by balancing cost, site restrictions, and performance. Given the constraints and insights from simulation, the design of HAWC ensures that it functions effectively as a  calorimeter for secondary electromagnetic particles. The water depth above the PMTs ( $\sim$ 4 m $\approx$ 10 radiation lengths) optimizes sensitivity to Cherenkov emission from secondary electromagnetic particles. 

If the water column above the PMT is too large, the sensitivity to the number of photoelectrons (PEs) per energy will be too low. If the water above the PMTs is too shallow, the response due to the proximity of the secondary electromagnetic particles from the EAS to the photocathode would be too large and not proportional to the deposited energy. For the chosen  water depth, HAWC detects $\sim$ 40 PEs/GeV of deposited energy\footnote{This value is obtained by Monte Carlo simulations considering the characteristics of the PMTs, an energy loss of 2 MeV/cm, and generation of Cherenkov light of 390 photons/cm of track length between 300 nm and 700 nm.} for electromagnetic particles. More details of the configuration and detector design are presented in \cite {Sinnis2021,Smith2009,DeYoung2012,Abeysekara2013,delaFuente2013,Mostafa2014,Smith2016,Springer2016,Pretz2016,Hampel2016,Hampel2017,ICRC2015} and references therein. 

The use of individual tanks (as opposed to the large single water volume of Milagro) enabled HAWC to be constructed using a staged approach, which allowed for operating with a subset of the final array. We used these first observations to perform system-wide testing of components and the data acquisition system, ensuring a smooth transition to full operations. Even with a fraction of the complete array, HAWC was the most sensitive all-sky observatory in the VHE regime. The total time to complete construction was 3.5 years, beginning in 2011. HAWC-30 (with 30 WCDs), completed in September 2012, was used mainly as an engineering array. As we installed additional WCDs, they were incorporated into the data acquisition system. HAWC-111 operated from August 2, 2013, through July 8, 2014, with three to five times greater sensitivity than Milagro (\cite{ICRC2015}). The complete 300 WCD (HAWC) array began operations on March 20, 2015. More details about HAWC construction and its engineering prototype Verification and Assessment Measurement of Observatory Subsystem (VAMOS) are presented in \cite{Abeysekara2015a}.

HAWC has detected over 100 sources of VHE gamma rays and has contributed  significantly to our comprehension of the high-energy universe. The third HAWC catalog of very-high-energy gamma-ray sources is presented in \citep{Albert2020g}. The science topics (with representative references) can be summarized as:

\begin{itemize}

\item Discovery and studies of the TeV gamma-ray sky, including extragalactic sources such as active galaxies \citep{Albert2021a}, gamma-ray bursts and transients \citep{Albert2022GRB}, as well as Galactic gamma-ray sources such as PeVatrons \citep{Abeysekara2020,Albert2020a,Abeysekara2021a}, pulsar-wind nebulae and the Crab \citep{Albert2021pwn}, TeV Halo Objects \citep{Abeysekara2017a}, microquasars and binary systems \citep{Abeysekara2018a}, Fermi bubbles \citep{Abeysekara2017e}, and molecular clouds \citep{Albert2021e}.

\item Cosmic-ray studies of the anisotropy of arrival directions, all-particle energy spectrum, and composition studies \citep{Abeysekara2019c}.

\item Fundamental physics such as probing beyond the standard model of particle physics by searching for dark matter \citep{Albert2020c} and violations of the Lorentz invariance \citep{Albert2020d} in the high energy regime.

\end{itemize}

Other topics include searches for primordial black holes evaporation \citep{Albert2020e} and studies of the Sun, the interplanetary medium \citep{Akiyama2020} and space weather \citep{Alvarez2021}. The synergy of HAWC with other observatories such as IceCube, VERITAS, \textit{Fermi}-LAT, and H.E.S.S. is exemplified in \citep{Ayala2021}. 

\section{Components of the HAWC Detector Hardware}
\label{S:site}
We describe the steel tanks, bladders, water, electrical infrastructure, cabling, and photomultiplier tubes that comprise the hardware elements of the HAWC observatory in the following sub-sections.

\subsection{Steel Water Tanks}
\label{sec:3.1}

Each of the 300 water tanks is made of corrugated, hot-dipped galvanized steel panels which are bolted together. The WCDs are covered with a dome-shaped, UV-hardened, sand-colored, scrimmed, vinyl polyester fabric roof to prevent rain and snow accumulation on the bladder and to serve as an additional light barrier. The ground inside the steel structure is flattened and covered with a layer of sand to prevent rocks from penetrating the bladder. Geo-textile felt on the ground and walls prevents sharp objects such as bolts from puncturing the bladder.

We constructed the steel tanks top-down. We assembled the top ring of steel panels first, at ground level, and then raised it with jacks to allow the assembly of the next ring of panels. Then we raised these two rings again to allow another ring to be added below. This procedure continued until the bottom ring.  Finally, we lowered the entire tank into a .6m trench and back-filled it with rammed earth, serving as an anchor that prevents tank upheaval during an earthquake (HAWC tanks met 2011 standards for International Building Code Risk Category III~\citep{risk} and Seismic Design Category D~\citep{design}). Thus, we assembled the tank safely from the ground without working on ladders, scaffolding or a crane.

\subsection{Tank Bladder}
\label{bladder}

\begin{figure}[htb!]
    \centering\includegraphics[width=\textwidth]{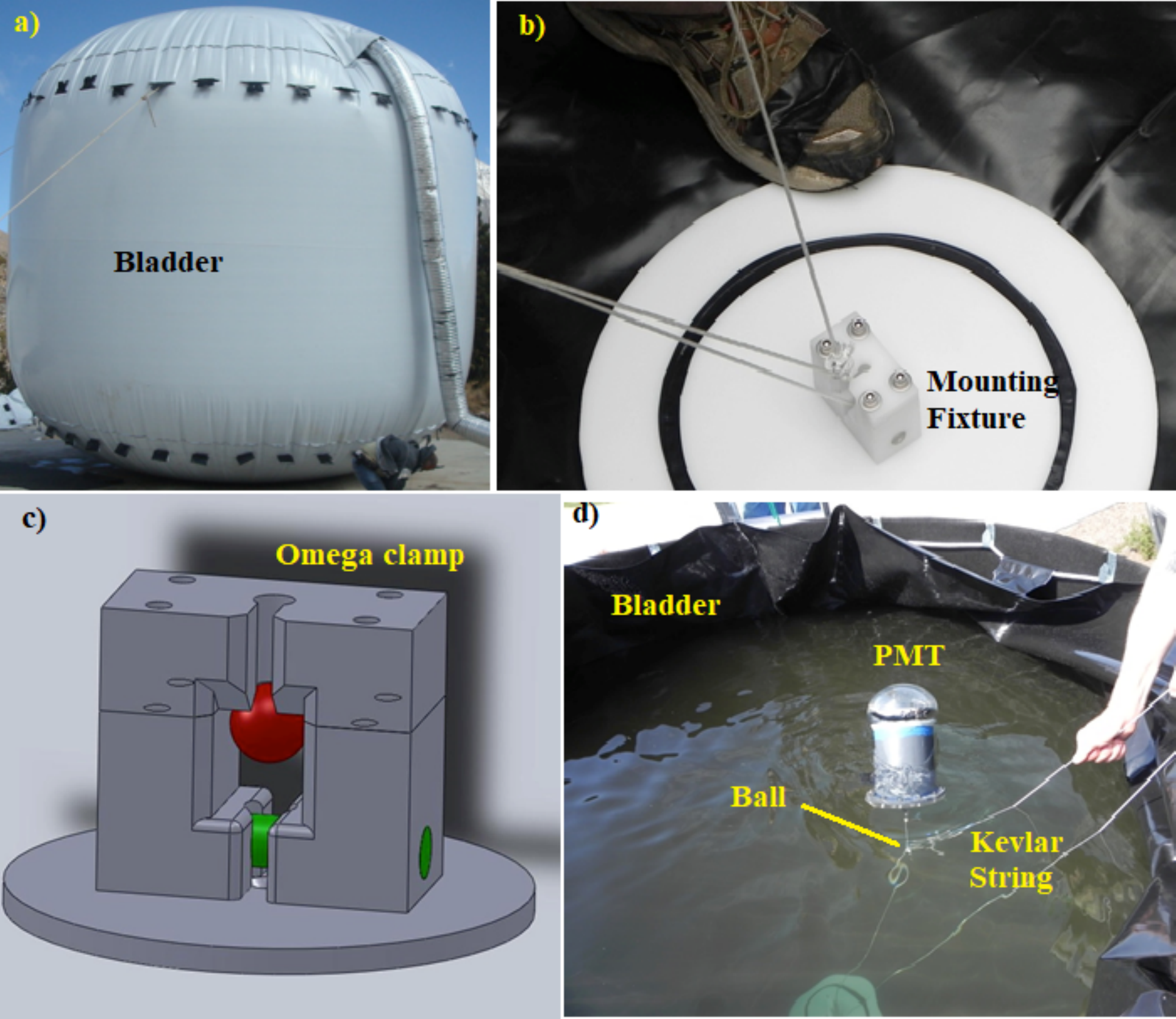}

\caption{ a) A bladder inflated to test for holes and light leaks. b) The Kevlar string and pulley system for a PMT. The mounting fixture is at the bottom of the WCD.  One end of the Kevlar string loop runs through the omega clamp (see c) and the other end attaches to the bladder hatch. c) Drawing of the omega clamp attached to the mounting fixture, which locks the acrylic ball from the PMT to the mounting fixture. d) The pulley system before installing a PMT in a WCD. The kevlar string, ball, and mounting fixture are labeled.}

\label{fig:bladder}
\end{figure}

The cylindrical tank bladders are made of flexible, low-density polyethylene. This is shown in Fig.~\ref{fig:bladder}a. Bladders enclose the water volume and act as a light barrier. The bladder material ($\sim$ 0.4 mm thick) is composed of two layers of three-substrate film fused/bonded during a co-extrusion process. 
 
Although the metal tank sides also provide a light barrier, the bladders act as the primary protection against external light reaching the PMTs. We tested to ensure that the laminate and seals were entirely opaque at single-photon levels for light with wavelengths between 260 nm and 600 nm. In addition, over the top of the bladder, we put a black film (agriculture foil) as an extra light blocker, covering the excess fiber to avoid light leaks. Each bladder has a PVC hatch on the top for access and installation of the PMTs and their cabling. The hatch has a film cover over it to block light passage through the hatch and the penetrations of the cables and optical fibers. The bottom of the bladder has an integrated mounting fixture for each PMT.

The part of each mount outside the bladder bottom is attached to a stake in the ground, which was surveyed prior to tank construction, to ensure that the position of the PMT is known to be within 0.16 cm.

In Fig. 3 b, c, and d, we show the pulley system to install the PMTs after the tanks are filled with water. This system decouples PMT deployment from water delivery and allows easy replacement of PMTs during operations. A loop of Kevlar string connects the hatch of the bladder (at the top of the tanks) with the PMT mounts at the bottom of the bladders. The bottom of each PMT is attached to a small plastic ball which locks the PMT to the mounting fixture. The PMTs point upward due to their buoyancy of about 80 N. PMTs can be easily removed by pulling from the other side of the Kevlar loop and disengaging the ball from the mount.

\subsection{Water Filtration System}
\label{sec:3.2}

We soften, sterilize, and filter well and surface water to fill the WCDs.  HAWC water filtration is a multi-step process to provide clear, pure water, that starts at the water source and continues at the HAWC site. 

At the well (located in the valley at 2640 m.a.s.l.), the water goes through a 30-micron pre-filter before it enters the softener system, where the hardness is reduced to below 2 grains. We do the softening at the well because the softener backwash water cannot be dealt with at the HAWC site since it is in a national park. Trucks of 15 kiloliter capacity transport water from the well to HAWC.  At HAWC, the water is pre-filtered again at 15 microns before it is transferred to one of five ``dirty'' water storage tanks.

\begin{figure}[htb!]
\centering
\includegraphics[width=\textwidth]{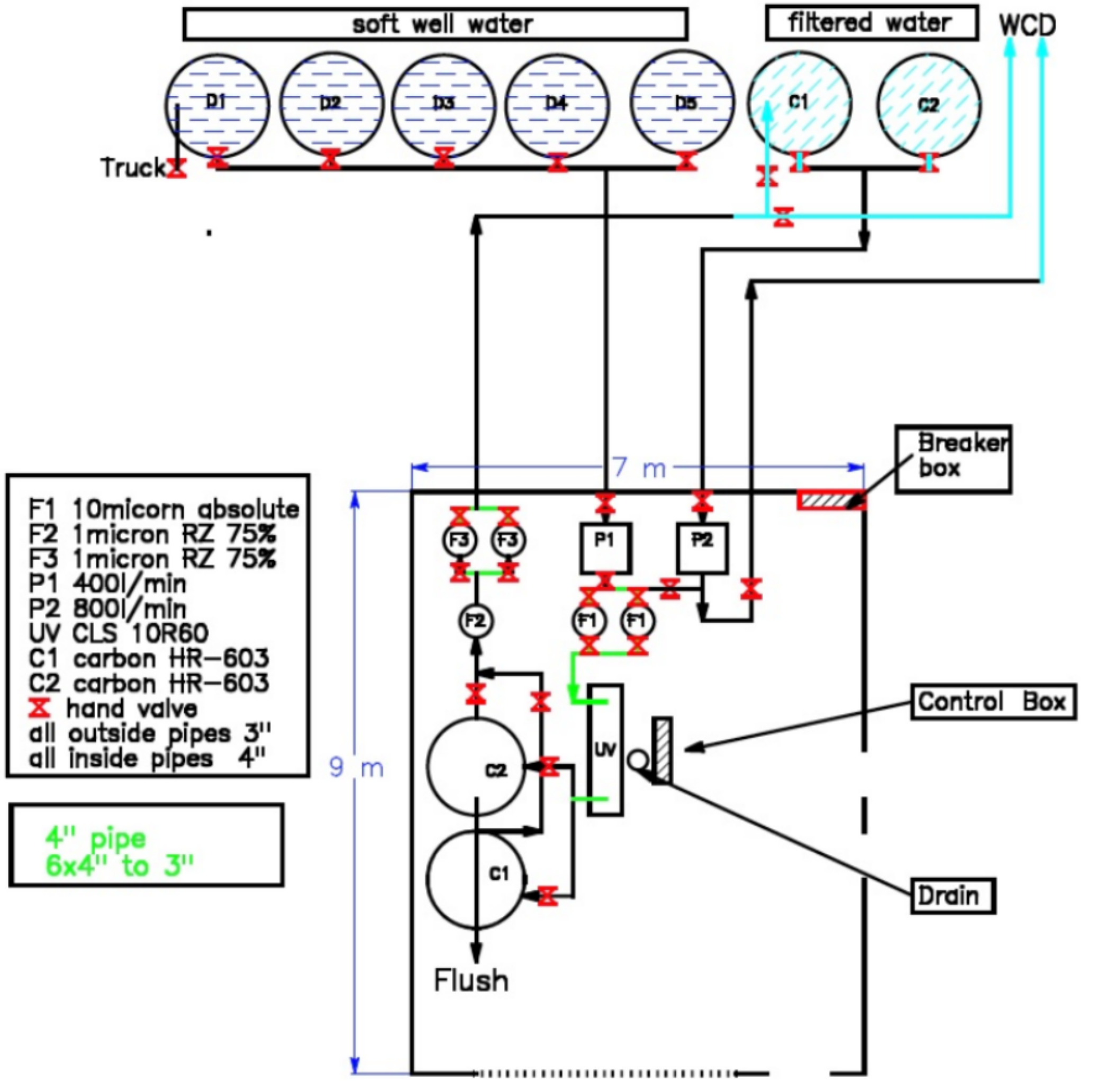}
\caption{Scheme of the water filtration system}
\label{fig:figWfs}
\end{figure}

The filtration system is designed to run 8 hrs/day and can fill at least one tank  (13 trucks) per workday. Trucks ran up to 7 days a week when recovering from bad weather or transportation problems.

Figure~\ref{fig:figWfs} shows the filtration system components, including one-micron filters, an Aquafine Model TSG 253 UV sterilizer \cite{UV}, charcoal filters, and a second set of one-micron filters. The throughput is limited by the internal impedance and the flow capacity of the UV sterilizer.

The filtered water is stored in two ``clean'' HAWC water tanks. A pump connected to a fixed underground PVC pipe and flex pipes fills a WCD in seven hours. If water needs further filtration, it can be pumped back through the filtration system.

\subsection{Attenuation length and water transparency}
\label{sec:3.3}

The attenuation length $\lambda$ describes the water transparency and is computed from the light-path length $l$ and fractional transparency $f$ by inverting $f = \exp(-l/\lambda)$. A sufficiently large attenuation length is necessary for the successful long-term operation of a WCD, to avoid loss of Cherenkov light due to the presence of organic material (e.g., bacteria and algae). Our instruments measure the fraction of light arriving at a photodetector, which is dominated by apparent losses due to scattering. For detector purposes, scattering matters less than actual light absorption, but our measurement of the combined scattering and attenuation length is sufficient to monitor relative water quality. We chose a minimum $\lambda$ of 15 m in the filtration system and 10 m inside the WCD. We determined these values by considering the 4.5 m water column inside the WCD,  the sensitivity of PMTs, and the duration of HAWC operation ($\sim$ 10+ years). 

To monitor the transparency and water quality of HAWC, we perform several measurements both on-site and in a laboratory. We obtain these water samples from: water collection at the HAWC site (melting snow and natural springs), the truck delivering the water from the well, after the water filtering processes, when filling each WCD with water, and from inside the WCDs over time. 

We use a convenient commercial transmission meter (C--star \cite{star}) for water quality monitoring. Typical attenuation measurements do not consider the UV regime.  The transmission meter operates with a 470 nm blue light laser diode as the emitter, which travels a path length of 0.25 m until it reaches the receiver. In each measurement, the fractional  transparency f$_r$ is determined as \citep{eq1}:

    \begin{equation}
    \label{eqn:eqn1}
    f = \frac{C_{sample} - C_{darkness} }{C_{calibration} - C_{darkness}},    
    \end{equation}
    
where C$_{sample}$ is the measurement for the sample, C$_{darkness}$ is the respective dark current, and C$_{calibration}$ is the measurement for distilled water. The transmission length is determined by

    \begin{equation}
    \label{eqn:eqn2}
    \lambda = \frac{-l}{\ln(f)}.
    \end{equation}

We also performed more precise calibration measurements in a reference laboratory (see \cite{Garfias2012} for details). That setup consisted of a 30 mW power 405 nm wavelength laser, a spectrometer, and a longer 1 m cylindrical tube with two quartz windows. Using an empty, dry, and clean tube, we measured the loss per quartz window to be between 5\% and 7\%. Measurements for the transmittance of the reference distilled water are taken with the same tube and the same orientation. With this system, the corrected fractional transparency $f'$ for the water alone is determined by:

\begin{equation}
\label{eqn:eqn3}
   f' = \frac{C}{A} + f_w,
    \end{equation}
    
where C is the measurement of the water sample, A is  the intensity of the laser, and $f_w$ is the fractional loss per window (with water in the tube, only the two air/quartz interfaces are important). 

This transmission measurement for the water alone allows comparison of the C--star measurements with the measurements performed at the laboratory, and confirms that the transparency of water reaches at least 15 m in the filtration system and about 12 m in the WCDs.

To save money on water transport from the well, we used some water collected from melting snow and natural springs.  This water (especially when spring flow is low) has a yellowish color due to lichens, which remains even after all organic matter is filtered out.  Processing the discolored water depleted the charcoal and stained the filters. We diluted all spring water with well water, and eventually stopped using it. Some tanks with mixed water had scattering/attenuation length as low as 8m.  Less than a quarter of the tanks had $\lambda$ of 8-10 m; the vertical muon charge deposition in such tanks was less than 3\% different from that in tanks with $\lambda > $ 10m.

\subsection{AC Power and Backup Solar Photovoltaic System}
\label{sec:3.4}

The Mexican power grid is HAWC's principal electrical power source. We extended the existing medium high voltage (HV) transmission line (installed for the LMT) by one km to the HAWC site. A 225 kVA transformer steps down the 34.5 KV transmission line to 3-phase 220 V for distribution to equipment throughout the HAWC site. 

HAWC uses backup solar power for essential services. The HAWC 4.1kW solar-power system has 18 panels. A deep cycle battery bank, which in normal conditions is charged by the AC line voltage, or in its absence by the solar panels, provides at least a week of power for critical monitoring and communication tasks. 

\subsection{Electrical AC grounding}
\label{ACgrounding}

HAWC is constructed on the slope of a volcano that has low soil conductivity. The ground impedance is typically around a hundred Ohms measured with a four-point ground monitoring technique and not suitable for a safe electrical ground without special treatment. Therefore, HAWC has three separate electrical grounds to provide a secure electrical system. The first electrical earthing is at the central transformer station where the incoming underground AC high voltage is transformed into three-phase 220 V and 120 V. A second earthing is at the HUB, which is created using the rebar iron in the large concrete pad in addition to long copper stakes --  a system referred to as a Ufer ground or a concrete-encased electrode (CEE) by NEC guidelines \citep{nec,necwiki}. The third and most critical earthing is established at the CH substation for electronics. This transformer provides power for all HAWC electronics through 2 UPS (Uninterruptible Power Supply) systems.  The 4 m $\times$ 10 m ground structure is based on three components: 3 m long copper stakes, large area short copper stakes to increase surface contact, and ground enhancer along the connecting copper wire. This system is about 0.5 m underground to reduce effects of surface drying. Earth resistivity measurements show a typical value of 2.5 $\Omega$ to 5.0 $\Omega$ depending on the season. This ground is also used by the outdoor spark gaps which protect every copper cable entering the CH against voltage surges; optical fiber connections are used wherever possible.

\subsection{Lightning and Grounding Improvements}
\label{sec:3.6}

In 2014 lightning struck about 200 m from HAWC and damaged electronics.  Although lightning is common at HAWC in the summer, most strikes land nearer the peak of Sierra Negra (close to the LMT). Most of the damage was to the emitter-coupled logic (ECL) interfaces among electronics components, requiring expensive and lengthy repair. Parts of the water-level monitoring system were also impacted. The damage was likely caused by different electronics crates floating to different levels, perhaps 10V or more apart, exceeding the allowed range of differential ECL inputs. The damage mechanism was not completely understood. 

We examined HAWC grounding inside and outside the CH and took many preventive actions:

\begin{itemize}

\item We ensured by measurement, solid connections of each VME crate to the ground, adding jumpers as needed.

\item We tied crate grounds to each other more strongly with copper braid to the central point (the transform AC ground), and we connected them to the racks with serrated washers.  We also added (for fast transients) copper foil/tape amid crates of a rack, among racks, and below cable, paths running between racks.

\item We changed the GPS signal entry to the CH from a wire cable to a radio repeater.

\item We separated solar power from ``clean'' electronics power and grounding.

\item In addition, we connected all the steel WCDs to each other in a large mesh made  with 6 AWG (American Wire Gauge \citep{awg}) copper wire. This eliminates surface voltage differences between the WCD structures which had previously been observed when low clouds passed over the array. This WCD grounding is not connected to any electrical ground and has no electrical connection to PMTs inside tanks.  This change did not, as hoped, much reduce PMT rate changes due to local electric fields.

\end{itemize}

We had operated for 2-3 years before the problem occurred. We have not had lightning damage in the CH for eight years after these preventive actions were taken, despite significant lightning storms and strikes within 20m of the HAWC main array.

\subsection{Cabling}
\label{sec:3.7}

\begin{figure}[htb!]
\centering\includegraphics[angle=90,width=\textwidth]{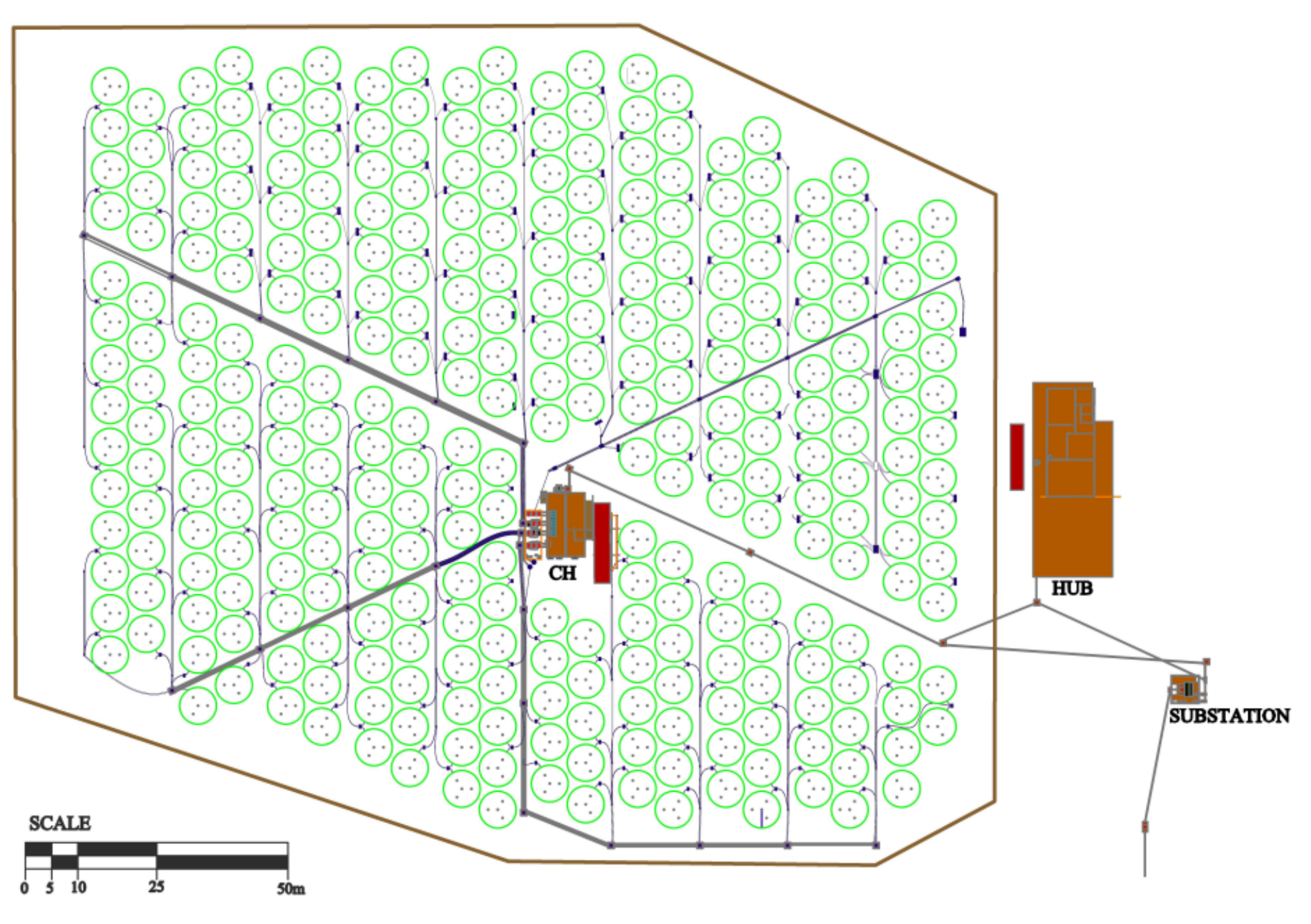}
\caption{Cabling layout (grey-violet solid lines). The HAWC platform is outlined in brown. Power lines are shown running from the substation to the CH and the HUB. The spark gaps are located next to the CH, where all cabling converges.  }
\label{fig:cabling}
\end{figure}

Several hundred km of cables and optical fibers (installed about 1m underground) connect to the WCDs of the HAWC observatory. From each WCD, four HV/signal cables (one per PMT), one optical fiber, and one ethernet category five (CAT5) cable run to the CH. 
Each PMT base is directly connected to an SHV-terminated cable running to a tank-side access box in which these cables connect to the long HV cables that run toward the CH. Junction boxes for the optical fibers and CAT 5 cables are also mounted on each tank. Locating the CH at the center of the array minimizes the cable length to the most distant PMTs. Figure \ref{fig:cabling} shows the layout of WCDs and cable trenches. 

\subsubsection{High-Voltage Cables and their Spark Gap Protection}

HAWC HV/signal cables are RG-59 Belden 8241 coaxial cables designed for analog video signals.  They have an external diameter of 6.15 mm, and a central wire gauge of 23 AWG \cite{belden}. This cable type was successfully used (for outrigger detectors) in Milagro. The cables can withstand 3100 V DC, so they are appropriate for carrying both HV and our PMT signals together on the same cable. The PMT signal is isolated from the DC HV through a pick-off capacitor in the HV sector of the front-end electronics (Section \ref{S:Electronics}).  

\begin{figure}[htb!]
\centering\includegraphics[width=0.5\columnwidth]{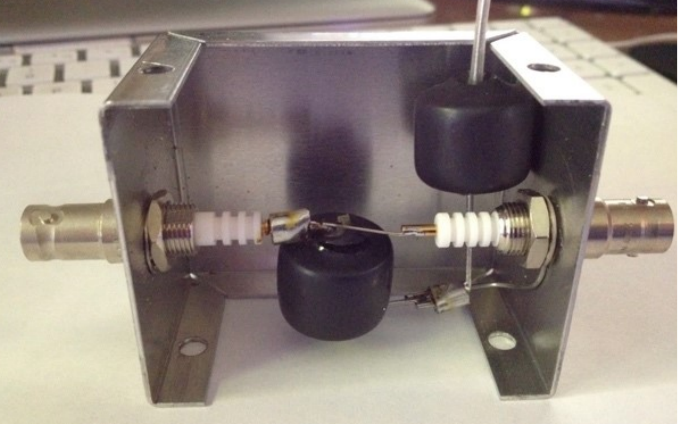}
\centering\includegraphics[width=0.5\columnwidth]{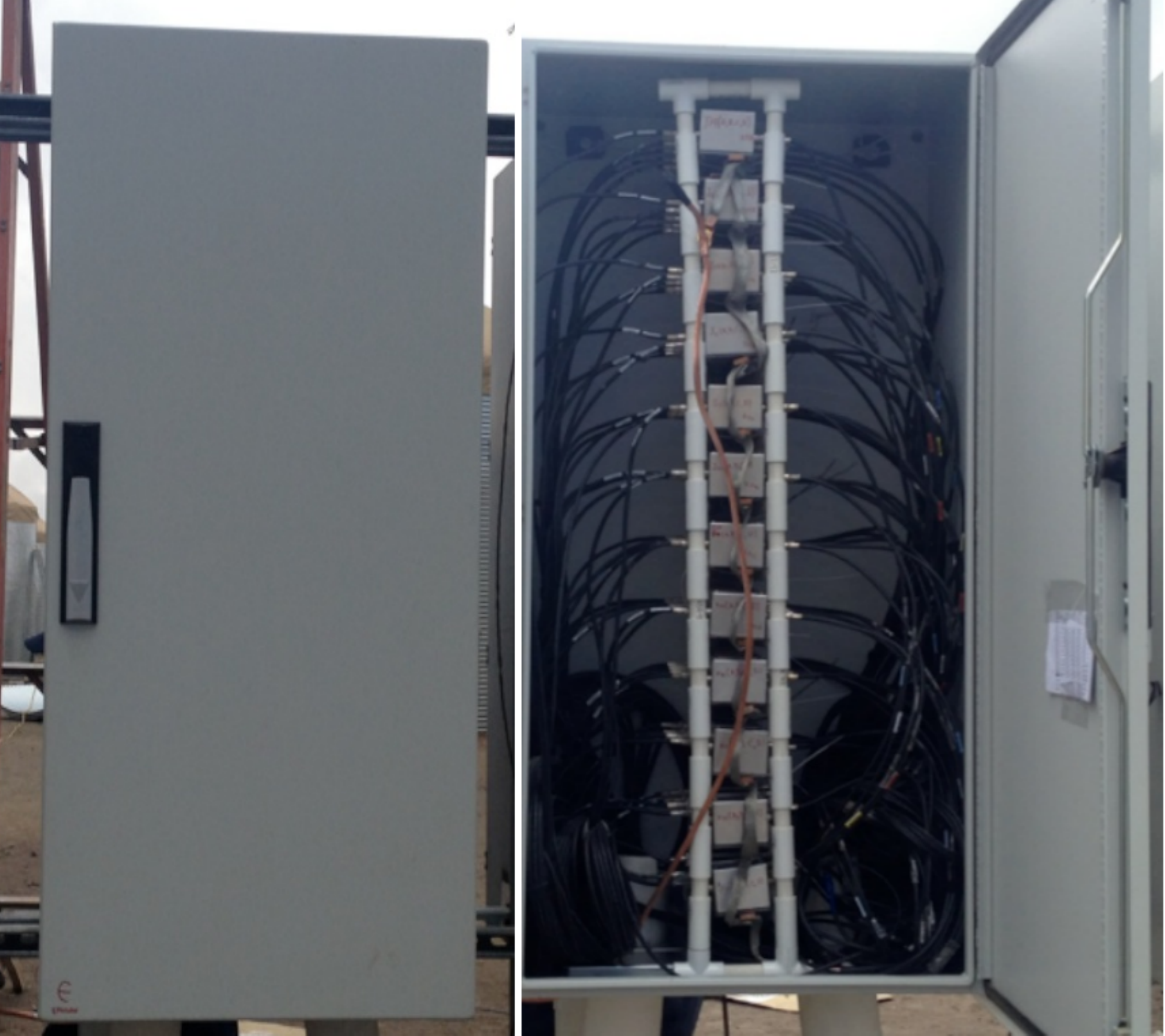}
\caption{\textbf{Top:} The spark gap protection device used for HAWC RG-59 signal/HV cables. On an over-voltage the spark-gaps clamp the conducting wire to the HAWC ground at the counting house. \textbf{Down}: A cabinet containing spark gap protection devices. }
\label{fig:spark_box}
\end{figure}

To ensure cables have identical electrical properties, we delay-match each HV cable to a specific standard reference {\it Golden Cable}. The standard Golden Cable has a length of 149.3 m with a delay of 761.2 ns. A sample of 585 cables used in HAWC had a measured average time delay of 761.2 ns with a standard deviation of 0.8 ns. A full description of the HV cable manufacture and testing is given in \citep{Adams2016}. 

Spark gaps protect the electronics from high voltage bursts induced by lightning, transmitted through the signal/HV cables.  Each HV cable from the PMTs {\it (long cable)} connects to a spark gap box in a electrical cabinet near the CH (see Figure~\ref{fig:spark_box}). From these spark-gap boxes, a 10 m length of RG-59 cable {\it (short cable)} runs to the CH. Buried boxes near the cabinets store any excess cable length of {\it long cables}, as well as of optical or CAT5 cables.  

The HAWC spark gap has the same design as used in Milagro.  The device is a small aluminum box (shown in Figure~\ref{fig:spark_box}) containing two Teledyne-Reynolds spark gaps and SHV connectors for the incoming and outgoing HV/signal cable. A DKF-3000L spark gap tube is connected between the conductor and the cable shielding braid, limiting their voltage difference to 3 kV.  A DFK-0230L spark gap between the cable shielding braid and the HAWC ground structure, limits their voltage difference to 230 V.

The spark gap tubes have an initial resistance of over a gigaohm, and a capacitance of less than 10 pF.  They do not interfere with the DC HV or PMT signal.  HV shrink tubing insulates the spark gaps.

The spark gaps are rated for 40,000 amperes (for a wave shape with an 8-microsecond rise and 20-microsecond decay to 50\%) and can sustain up to three to five lightning strikes. The aluminum box itself connects the input and output shield braids but is otherwise floating and does not connect to the ground structure or any other device. Individual boxes are grouped and mounted inside a weatherproof outdoor electrical cabinet. To prevent condensation during weather below the dew point, we added heating tape to warm each spark gap cabinet. 

\subsubsection{Fiber Optics}

The HAWC optical fiber system is used to measure the relative PMT time offsets vs. signal amplitude as well as to calibrate the signal amplitudes. The optical fibers run from the calibration system (see Section \ref{S:Calibration} ) fiber-optic distribution network in the CH to distribution boxes at the base of each pair of neighboring WCDs, where they connect with a local fiber leading to a diffuser ball in each WCD tank.

\subsubsection{Water Depth Monitoring Cables}

A 12m cable attached to the water depth sensor (see Section \ref{S:ems}) at the bottom of the tank ends in a junction box at the WCD with an XLR connector. Unshielded Twisted Pair (UTP) CAT5 cables run from junction box toward the CH.  These cables have eight conductors, sufficient for signals from 2 WCDs.  As with the signal/HV cables, the long CAT5 cable run goes to an electrical cabinet near the CH containing ethernet spark gaps, after which a 12 m cable runs to the CH.  

\begin{figure}[htb!]
\centering\includegraphics[width=0.5\columnwidth]{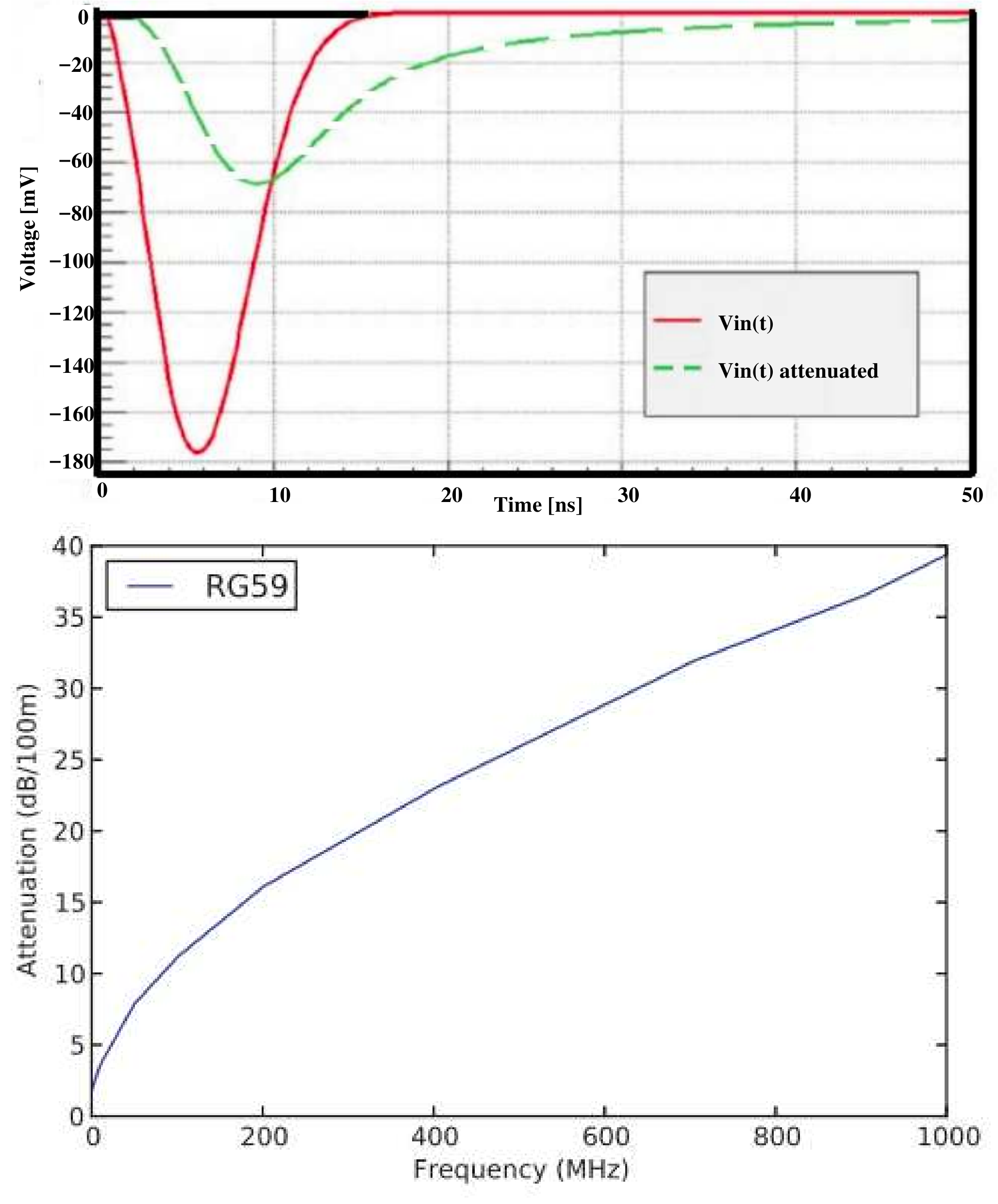}
\caption{\textbf{Top:} Simulated response for a PMT pulse with and without the cable attenuation. \textbf{down:} attenuation due to 100 meters of RG59 cable as a function of frequency from 1 MHz to 1 GHz is shown. This  frequency range is relevant for PMT pulses, and shows strong attenuation at the highest frequency values. Based on \citep{Wisher2016}.} 
\label{fig:PMT}
\end{figure}

\subsection{Photomultipler Tubes}
\label{sec:PMTs}

As mentioned in Section \ref{S:HMDandOp}, HAWC has two PMT types: Hamamatsu R5912 and R7081-02, which have the following specifications:

\begin{itemize}

    \item Spectral response from 300 to 650 nm with a peak response at a wavelength of 420 nm.
    \item Dynode chain of 10 stages
    \item Anode Dark current 100 nA (typical) 1000 nA (max)
    \item Typical Rise Time of 3.6 ns
    \item Typical Transit Time of 62 ns with a spread of 2.4 ns
 
\end{itemize} 

The R7801 has 530 cm$^{2}$ of photocathode area, $\sim 40$\% larger than the R5912 (340cm$^{2}$). The increase in area and efficiency of the R7081 enhances response, while only increasing the transit time spread from 2.4 ns to 3.4 ns.
More details are reported in the respective Hamamatsu manuals \cite{hamamatsu}.

The PMT base consists of a passive resistor chain with high voltage capacitors on the last two dynodes in the chain to prevent any voltage drop from large pulses. To maximize the sensitivity to single photoelectrons and compensate for the cable attenuation (see Figure \ref{fig:PMT}), we operate the PMTs a high gain of $\approx 10^7$. The dispersion in the cable is calibrated to minimize its impact on timing due to threshold discrimination electronics. The bases have generally been stable, but some carbon Megaohm HV resistors have degenerated over time (many of these bases are over 20 years old), causing slow or catastrophic decreases in gain.  HV capacitors (perhaps due to component quality issues) have also failed in some newer bases. We have an ongoing campaign to replace broken carbon resistors with metal film resistors. 

We measured the gain of each individual PMT and grouped PMTs with similar voltage versus gain curves into the same WCD tanks. There are  
 advantages to grouping similar PMTs in the same tank:

\begin{itemize}

    \item The high voltage cables for eight PMTs (a WCD tank pair) are connected to a single HV supply channel. Therefore, matched PMTs have similar gains.

    \item The PMT gain in all tanks can be made similar by selecting an appropriate voltage for each HV channel, simplifying data analysis and simulation.

\end{itemize}

We measured the dependence of response as a function of photon impact position  
of the R5912 PMTs using a Robotic Characterization System (RCS) \citep{Vanegas2016}. This system automatically measures PMT response at 101 locations distributed over the PMT's spherical active surface. A charge integrating ADC with a 20 ns time window and a conversion of 0.25 pC per channel digitized pulses from PMTs in response to the LED's flashing. We varied the PMT's HV from 1400 V to 1850 V to measure gain curves. With typical operating voltages corresponding to a gain of 1.5$\times$10$^7$, the charge spectra for the LED mean light output varied from 1 to 5 PE. A detailed description is available in \citep{Wisher2016}.  Subsequent measurements,  with varying light intensity, allowed us to distinguish gain variation from charge collection efficiency; we found that PMTs suffer significant collection efficiency loss at the edges the photocathodes.


\begin{figure}[htb!]
\centering\includegraphics[width=\columnwidth]{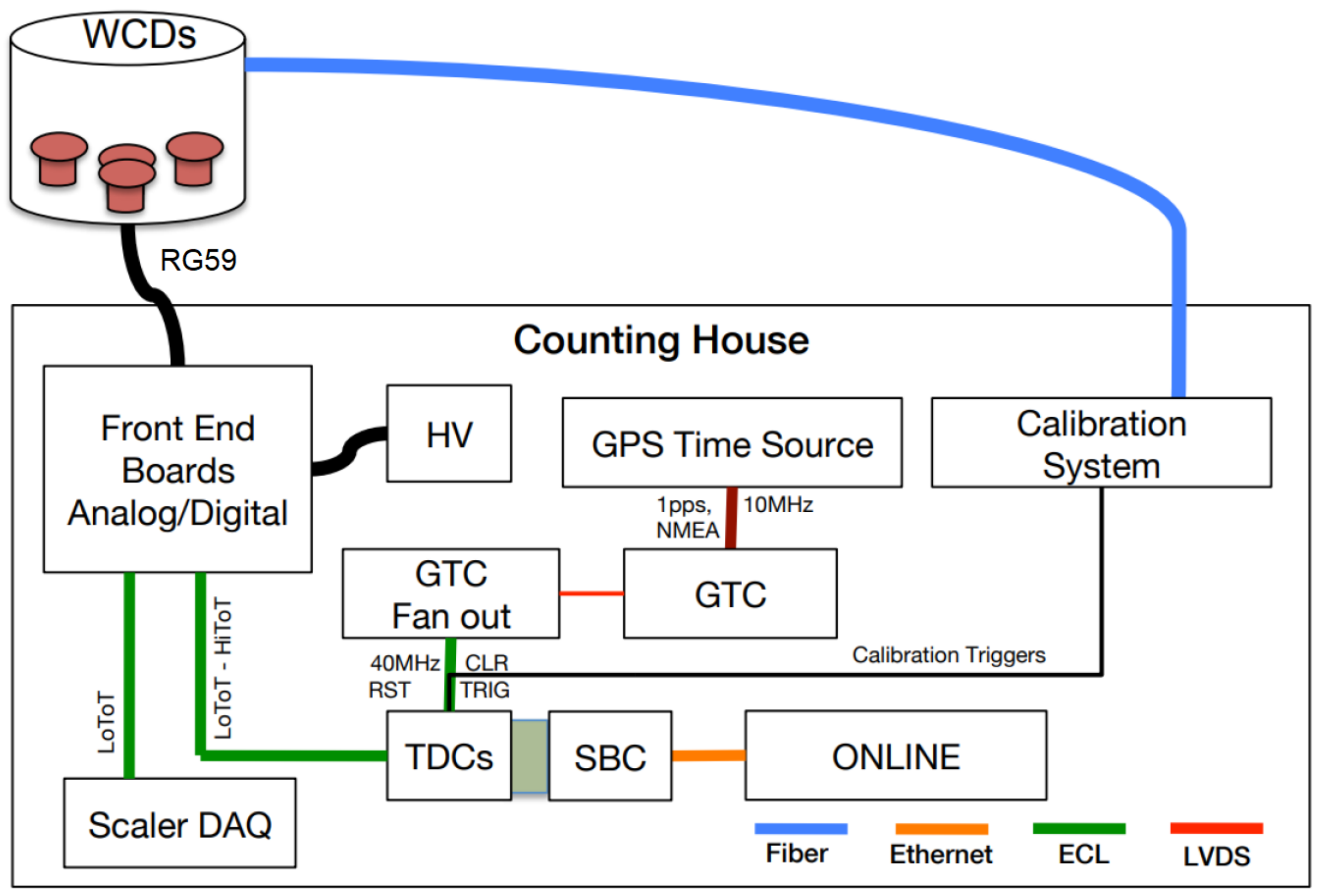}

\caption{Top-level diagram of the HAWC electronics showing a summary of the critical subsystems and the interconnections, including HV and optical fiber cabling. Based on \citep{Wisher2016}. NMEA refers to the National Marine Electric Association format in which GPS presents data \citep{gpsnmea,wikinmea}; CLR, TRG and RST are control signals for the TDC system.  The LoToT and HiToT time over threshold signals are discussed in section \ref{S:FEB}.
}
\label{fig:TLD}
\end{figure}

\begin{figure*}[htb!]
\centering\includegraphics[width=\columnwidth]{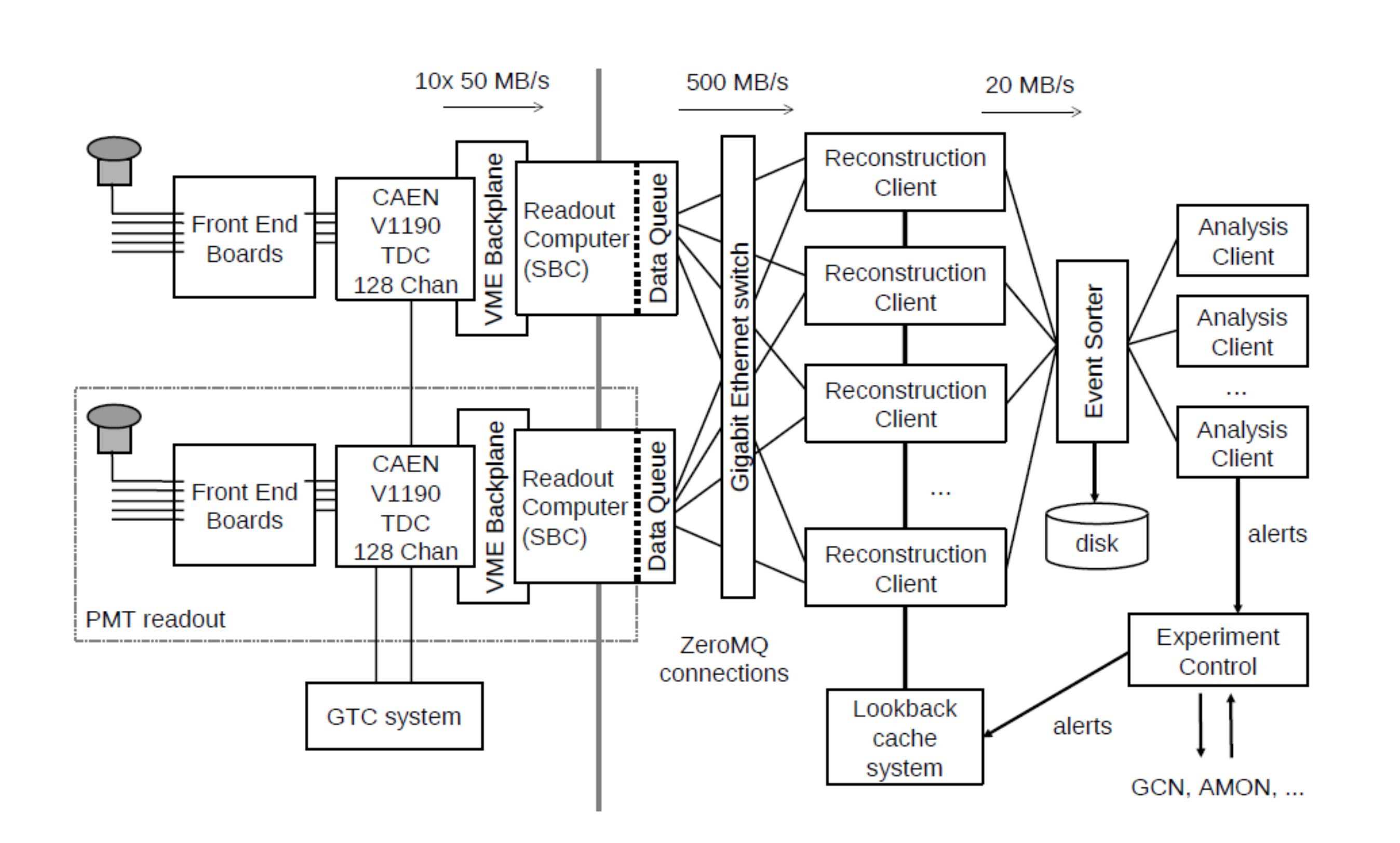}
\caption{Schematic overview \citep{Abeysekara2018f} of the HAWC data acquisition and online processing system, as described in the text of section \ref{S:Electronics}.  }
\label{fig:DAQ_overview}
\end{figure*}

\section{HAWC Readout Electronics} 
\label{S:Electronics}

Here we describe the components of the readout electronics chain.
Figure \ref{fig:TLD} shows a schematic overview of the HAWC electronics in the CH. Front-end boards (reused from Milagro) make digital signals from the PMT inputs, including a time-over-threshold (ToT) measurement of amplitude at two thresholds. Each PMT's ToT signal is recorded by a commercial Time to Digital Converter (TDC). A central custom GPS timing and control system (GTC) provides timing and control of all TDCs. Single-board computers (SBCs) read out the TDCs. The SBCs group data into fixed-length time blocks before sending them to the online reconstruction CPUs. A separate data acquisition system records hardware scaler rates. The main elements of HAWC electronics are described in the following sub-sections. Further details about the electronics can be found in \cite{Wisher2016}.  We found it very beneficial to develop and test software and hardware interfaces at a comfortable  integration site before shipping and installation in the HAWC CH.

Figure \ref{fig:DAQ_overview} shows a sketch of the HAWC DAQ system. The control computer starts and stops data taking runs and interacts with the GTC electronics to manage the TDCs and readout computers. The DAQ and online processing system assemble, store and reconstruct the TDC data. Analysis clients provide real-time event-based monitoring, and analysis clients can produce real-time alerts (see Section \ref{sec:real_time}). The DAQ system and software are discussed in more detail in \cite{Abeysekara2018f}.

\subsection{Front-End Board (FEB) Electronics}
\label{S:FEB}

\begin{figure*}
\centering\includegraphics[width=\textwidth]{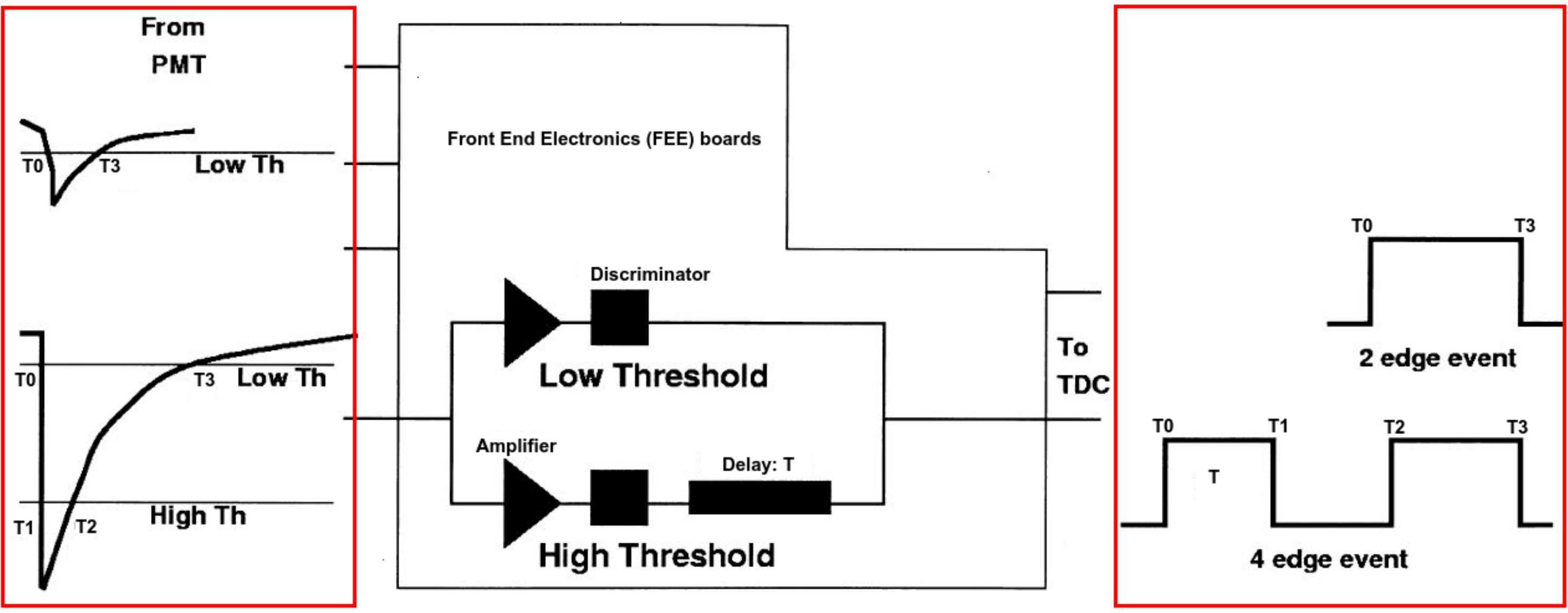}
\caption{The analog PMT signals are split and passed through two paths. In each path, there is an amplifier and discriminator circuit. The ratio of the amplifier gains is 7 to 1. The higher gain circuit has an effectively lower threshold (Low Th). There is a time (T) delay in the high threshold (High Th) path. The 2-edge event is related with the Low Th, while the 4 edge event is related to the High Th. 
}
\label{fig:feb_tot}
\end{figure*}

The cable from the PMT (after the spark gap) connects to a FEB. The signal cable supplies HV to the PMT and the analog anode PMT signal is AC-coupled to the RG59 cable. The PMT signal is capacitively picked off at the FEB,  amplified and shaped with a 75 ns time constant, then digitized by a two-threshold ToT circuit. The time at which the threshold is passed provides the timing used in the angular reconstruction. 

Figure~\ref{fig:feb_tot} shows PMT signals (on the left) and associated FEB discriminator outputs (on the right). For a small signal (upper waveforms on the left and right), the resulting output of the FEB is a single digital pulse with two edges at T$_0$ and T$_1$: the times the pulse crosses the Low Th threshold while rising and falling (respectively). The time duration that the PMT signal exceeds the low threshold, LoToT, in this case, is T$_1$--T$_0$.
For the large-signal case (lower waveforms), the output of the FEB is two digital pulses with four edges at T$_0$, T$_1$, T$_2$, and T$_3$. These are the times of rising crossings of the Low Th and High Th, and the falling crossing times of the High Th and then the Low Th. The time duration that the PMT signal exceeds the low threshold, LoToT, in this case, is T$_3$--T$_0$. The time duration that the PMT signal exceeds the high threshold, HiToT, is T$_2$--T$_1$. 

The FEB electronics are implemented in ECL logic which is fast but requires high power. FEBs are described in more detail in \cite{Atkins2000}. FEBs were modified in several ways for HAWC.  First, analog FEBs distribute two input HV levels to 8 PMTs, instead of one HV for 16 channels. Second, HAWC PMTs are operated at a lower gain than in Milagro to allow for larger showers without saturation.  As a result, the Low Th and High Th thresholds were reduced from 30 and 80 mV to 20 and 50 mV, maintaining thresholds at about 1/4-1/2 photoelectrons and 3-4 photoelectrons Low Th and High Th, respectively.  Finally, because the HAWC TDCs (see below) have better two-pulse resolution (5 ns) than Milagro TDCs, timing constants in the edge processing were adjusted to reduce the minimum delay time between Lo and Hi edges. This reduces the chance that two small hits nearby in time could be confused with a single large hit passing both Lo and Hi thresholds, since either scenario produces 4 edges.

\subsection{Power Supplies}
\label{S:PS}

Wiener PL506 low voltage power supplies provide the +5.2 V, +5 V, and -5 V for the FEBs.   Wiener 6023$\times$610 VME crates housed the scaler and TDC data acquisition system. 
A Wiener MPOD mainframe (with five ISEG\_EHS 32-channel 20125p modules) provides the PMT high voltage.  Selecting Wiener for all provides a consistent software control interface.  

The High Voltage passes through a custom breakout box to split the high-density 32 channel HV cable into individual SHV cables that connect to the front-end boards. PMTs were grouped to ensure that four tubes with consistent HV settings were deployed in each tank, and a single SHV connector fed two tanks with compatible HV requirements. Each front-end board holds 16 channels or four tanks.  This arrangement facilitated installation and repair, as only one HV channel has to be turned off to service a tank.  The PMTs are operated with positive HV so that the photocathode in the water is at ground.

\subsection{Scaler DAQ}
\label{S:Scalers}

The scalar DAQ provides a robust set of scaler rates, allowing the monitoring of rates even under severe weather conditions that overwhelm the TDC system, and independent of how we operate the TDC DAQ system. The hardware scaler system is also used in cosmic ray and solar physics analyses (e.g \cite{Alvarez2021} and references therein).
We use Struck SIS3820 VME scalers with rear transition modules with a density of 64 channels per VME slot. One Wiener VME crate records all 1200 PMT signals. Input signals come from the Low Th threshold output of digital FEBs for each PMT, and the monitoring system records rates once per minute.

\subsection{TDC DAQ}
\label{S:TDC}

The TDC DAQ for the HAWC PMTs is based on the 128 channel VME CAEN  V1190S-2eSST TDC.  These TDCs have  multiple-hit capability, record times with a granularity of .1 ns 
 and support several readout modes. As discussed in Section \ref{S:FEB}, the input to the TDCs combines arrival time and pulse height information, allowing measurement of ToT for two separate thresholds (a 2-bit nonlinear ADC encoding), whose analysis is described further in Section \ref{S:Calibration}.   

TDCs are located in two VME crates with each backplane subdivided into five independent sub-backplanes.  Each independent 4-slot sub-backplane holds a TDC and a GEXVB602 single-board computer by GE/FANUC (SBC) powered by an Intel i7 processor.  One SBC reads out each TDC using the 2eSST protocol \cite{sst} using a CENTOS5 library kindly provided by Sergey Boiarinov of the Jefferson Laboratory. The ten TDCs read out in parallel at 50 MB/s using this protocol, with a total data volume of 500 MB/s passing through a network switch to the online farm.  Rather than selectively reading out TDCs on air shower events, TDCs are periodically read out by a TRG signal generated every 25 $\mu$s, and the entire data stream is transferred to the processing farm. Each data frame covers a 26 $\mu$s time interval, slightly overlapping to avoid complex processing at frame boundaries.  The online farm applies a software event triggering criterion of 28 of the 1200 PMTs having hits within 150 ns and saves those hits; this multiplicity threshold is comfortably lower than the requirements of most physics analyses.  The multiplicity criterion fires at a 25 kHz event rate and reduces the stored raw data stream to 20 MB/s. Events are fully reconstructed online for real-time analysis and monitoring, but the full 20MB/s is saved for offline reconstruction.  The DAQ system is described in more detail in \cite{Abeysekara2018f}.

The HAWC PMTs exhibit some after-pulses due to ionized residual gas molecules inside the PMT which are accelerated to the photocathode. We veto after-pulses arriving several microseconds after a large hit, so they are not used in the analysis. This and other TDC-specific effects produce an estimated inefficiency of 1-2\% level (per PMT channel, not per event). 

Pre-pulses also occur but are mitigated by using the High threshold start time as a reference (whenever it is available) rather than the Low threshold start time.
 
In this high-throughput mode, the system has approximately a 2-3\% event deadtime fraction, as deduced by fitting the distribution of time differences between events.

\subsection{Timing and Control}
\label{S:GTC}

The primary responsibility of the HAWC GPS Timing and Control (GTC) system is to provide the control signals needed by the TDCs and scalers. The TDC design is based on a 40MHz clock. The GTC generates a 40MHz clock from the 10 MHz clock signal provided by the GPS. The GTC supplies TDC control signals which  zero counters and clears TDC event data; the signals arrive well separated from the 40 MHz clock edges.
These control signals are a crucial part of cleanly starting a run simultaneously in all TDCs and providing the internal event counts, and the counters of the 40MHz clock which label the TDC data headers. The GTC system was capable of handling either an asynchronous event-based trigger or a synchronous trigger.  HAWC chose the latter option, with the GTC providing a precise 40 kHz periodic TRG signal to the TDCs derived from the 40MHz clock.  The GTC also provides a periodic readout signal to the scaler DAQ crate.

The GTC encoded a sub-microsecond global time stamp for sky positioning in the TDC data channels. However, the analysis requirements were sufficiently met with NTP (Network Time Protocol) computer timestamps with an accuracy of approximately one millisecond. The GTC operation is described in a previous publication \cite{Abeysekara2018f} on the overall HAWC data acquisition system. The design of the GTC is discussed in detail in \cite{Abeysekara2014c}.

In 2018, we replaced the GPS timing portion of the GTC with a White Rabbit (WR) system with a custom WR-ZEN module \cite{WR} to synchronize the main HAWC DAQ with the HAWC outrigger DAQ, which has native timing based on the WR. WR implements the IEEE 1588 Precision Time Protocol (see e.g. \cite{1588}) in an open hardware framework. With the WR system, the 40 kHz readout TRG signal is synchronized to the GPS one pulse per second signal, and TDC runs are started at the top of a GPS second.  Therefore, the time of the event trigger is now determined to the accuracy of the GPS. HAWC's WR system will be discussed in greater detail in a future publication on the outrigger extension of HAWC.

\section{The Calibration of HAWC}
\label{S:Calibration}

Accurate reconstruction of the air shower requires precise timing and charge measurements from the PMT signals. To this end, we have a calibrated and monitored laser system that can deposit light in the tanks with known time and consistent amplitude.  We use the laser calibration system to determine in-situ PMT charge and timing characteristics. This calibration information is used to perform arrival time corrections of the particles from the EAS and to convert ToT to number of photoelectrons. The calibration system of HAWC, including charge, timing, and its analytical framework is summarized by \cite{Huntemeyer2009,Ayala2015,Younk2015}, with further details found in \cite{Zhou2015,Ayala2017} (and references therein). 

Two aspects of the calibration system are particularly worth emphasizing: we designed the hardware and software to operate remotely, and we record laser firing times in the TDC data stream so that we can take calibration data (a few hundred events/second) simultaneously with shower data, without causing significant down time. Below we describe the hardware of the laser calibration system and the analysis software.

\begin{figure}[!htb]
\centering\includegraphics[width=0.5\textwidth]{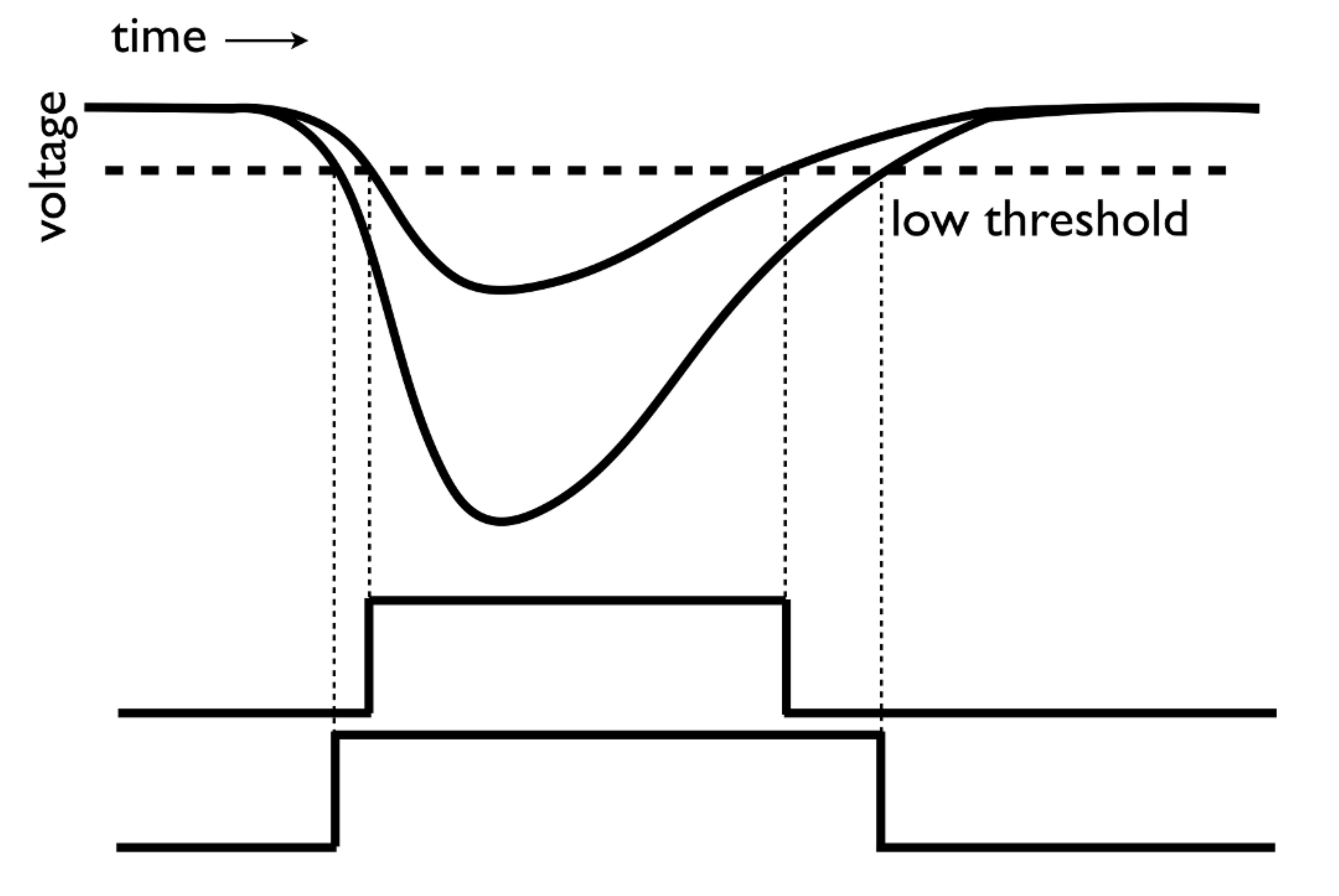}
\caption{Time slewing due to varying pulse height. As illustrated above, the discriminated pulse from a smaller signal starts later than a larger signal due to a delay in crossing the fixed voltage threshold (\citep{Zhou2015}).}
\label{fig:time-slewing-sketch}
\end{figure}

The charge and timing information for the energy deposited in WCDs of EAS are deduced from the width and leading-edge time of the discriminator pulses from each TDC (described in Section \ref{S:FEB}). Key tasks of the calibration system include determining the electronic time-slewing as a function of ToT and the offset between the PMT measured time and the fitted air shower front expected time (time pedestals) among PMT channels. Figure~\ref{fig:time-slewing-sketch} illustrates the time slewing effect resulting from a pulse leading edge crossing a fixed voltage threshold at a time which depends on the pulse amplitude.

\begin{figure}[!htb]
\centering
\includegraphics[width=\textwidth]{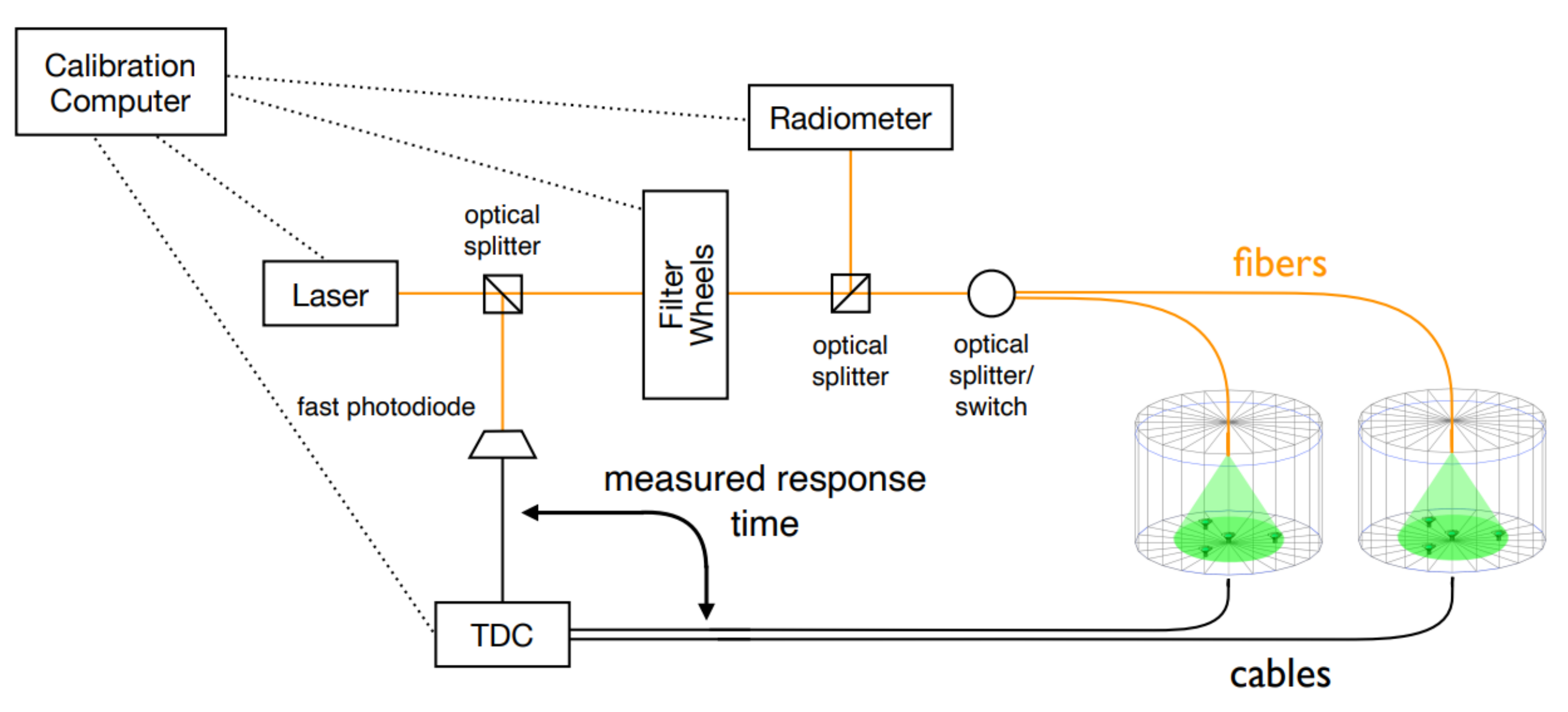}
\caption{A schematic of the laser calibration system \citep{Zhou2015, Hoskins2012}. See text for details.}
\label{fig:laser-system-scheme}
\end{figure}

Figure~\ref{fig:laser-system-scheme} displays the overall layout of the hardware for the laser calibration system. A laser-based optical system delivers short pulses of light to the WCDs. The PMT response to this calibrated light source is used to determine the relationship between ToT and PE for charge calibration. The measured time between laser firing (as measured by a photodiode and recorded by the TDC) and TDC edges is used to determine the PMT response time and cable delay, and measure the charge-dependent slewing effect and time pedestal for each PMT.

The optical system includes a Teem Photonics PNx-M green (532nm) laser~\citep{teem}, four LaserProbe RM-3700 radiometers ~\citep{laserprob}, optical splitting cubes, Spectral Products AB301 filter wheels~\citep{wheel}, 1:2, 1:4, 1:19, and 1:37 splitters, and chassis-based optical switches. Pairs of optical fibers of approximately 200 m length connect this system to and from each of the HAWC WCDs, as described in Section \ref{sec:3.7}. The laser delivers light in short pulses ($<$ 1ns). The fact that shower photons arrive over a longer period (up to 10ns, giving longer ToT for the same number of photons), which leads to one of the systematic uncertainties in HAWC calibration~\citep{Abeysekara2019a}.

Three filter wheels each hold six neutral density filters with different opacities to give 68 distinct filter combinations providing transmittance values ranging from 1 to 10$^{-6.5}$. Thorlabs optical diffusers, which hold the end of the optical fiber, are anchored to the bottom of each WCD tank and are submerged approximately 3 m above the central PMT. The  intensity of laser light at PMTs can vary from approximately 0.1 PE to more than 1000 PE. 

The relative intensity of the light before and after the filter wheel is recorded by the LaserProbe RM3700 Control Box radiometer \cite{laserprob} and a RjP-465 Energy Probe \cite{Rjp445}. At the lowest light levels, we determine the number of PEs by occupancy in each PMT, assuming Poisson statistics.  At higher light levels, we determine the number of PEs by scaling the occupancy-derived PEs by the ratio of radiometer values.  

The calibration computer controls the laser system devices ~\citep{Hoskins2012}. The regular TDC DAQ reads out the PMTs signals from the laser pulses. Two Thorlabs DET02AFC photodiodes provide calibration start and stop timing signals to the DAQ TDCs. The time the laser fired is recorded in a channel of the TDC DAQ system; the presence of the laser time in an event tags calibration events for later processing.

We do absolute charge calibration based on a physical standard: a clean sample of nearly vertical isolated muons passing though the center of the tanks.  We select tanks far from the shower core and with no hits in neighboring tanks and then require that the central PMT has a large signal and three outer PMTs hits occur at nearly the same time. We use the observed signals to measure the absolute sensitivity of each PMT and to provide an absolute normalization for the simulated PMT response. 

Timing corrections are done based on EAS data: we select clean air showers from nearly overhead with $>$ 200 hits and use average shower plane fit residuals to measure individual PMT timing offsets to within less than 1ns.  Then we incorporate these corrections into the calibration chain.

After these corrections, we use the charge information to determine the EAS core location and the lateral distribution of the EAS at HAWC.  We reconstruct the arrival direction of the EAS using the timing information. Combining all this information, we reconstruct the initial energy of the EAS.

The analysis software uses both the calibration system (radiometers) and the DAQ system (TDC data) to provide the final calibration products. These products are used in the reconstruction process to correct the different responses of the individual PMT channels. The calibration analysis software determines the charge calibration, which is the relationship between the measured ToT and the number of PEs detected by the PMT photocathode for varying light intensities. The relative time of the PMT responses and the slewing of measured times due to pulse-height variation is determined in the timing calibration.  As a final product, calibration constants are provided to the event reconstruction analysis system. The charge and timing calibration results and raw calibration data are archived in a dedicated database. 

\section{HAWC Operations}
\label{S:Operations}

We designed HAWC for autonomous operation.  The key to autonomous operation is comprehensive remote monitoring, automated control, and the absence of consumables.  The site is remote, and access has at times been restricted for scientists and even the local maintenance crew.  Down periods can become extended if spare equipment is not immediately available on site. The HAWC site has many features which motivate design for remote operation.  The volcanic soil is fine and abrasive, so we minimize traffic through the CH and clean it regularly.  That same soil makes electrical grounding difficult (see Section \ref{ACgrounding}). The site is subject to power outages due to electrical storms, heavy rain and snow, and even hurricanes and earthquakes.  Humidity varies widely at the HAWC site and can be low enough to require careful anti-static protections for handling electronics or high enough to require measures to prevent unwanted condensation.

\begin{figure}[!htb]
\centering\includegraphics[trim={.1cm 0 .1cm .1cm},clip, width=0.4\textwidth]{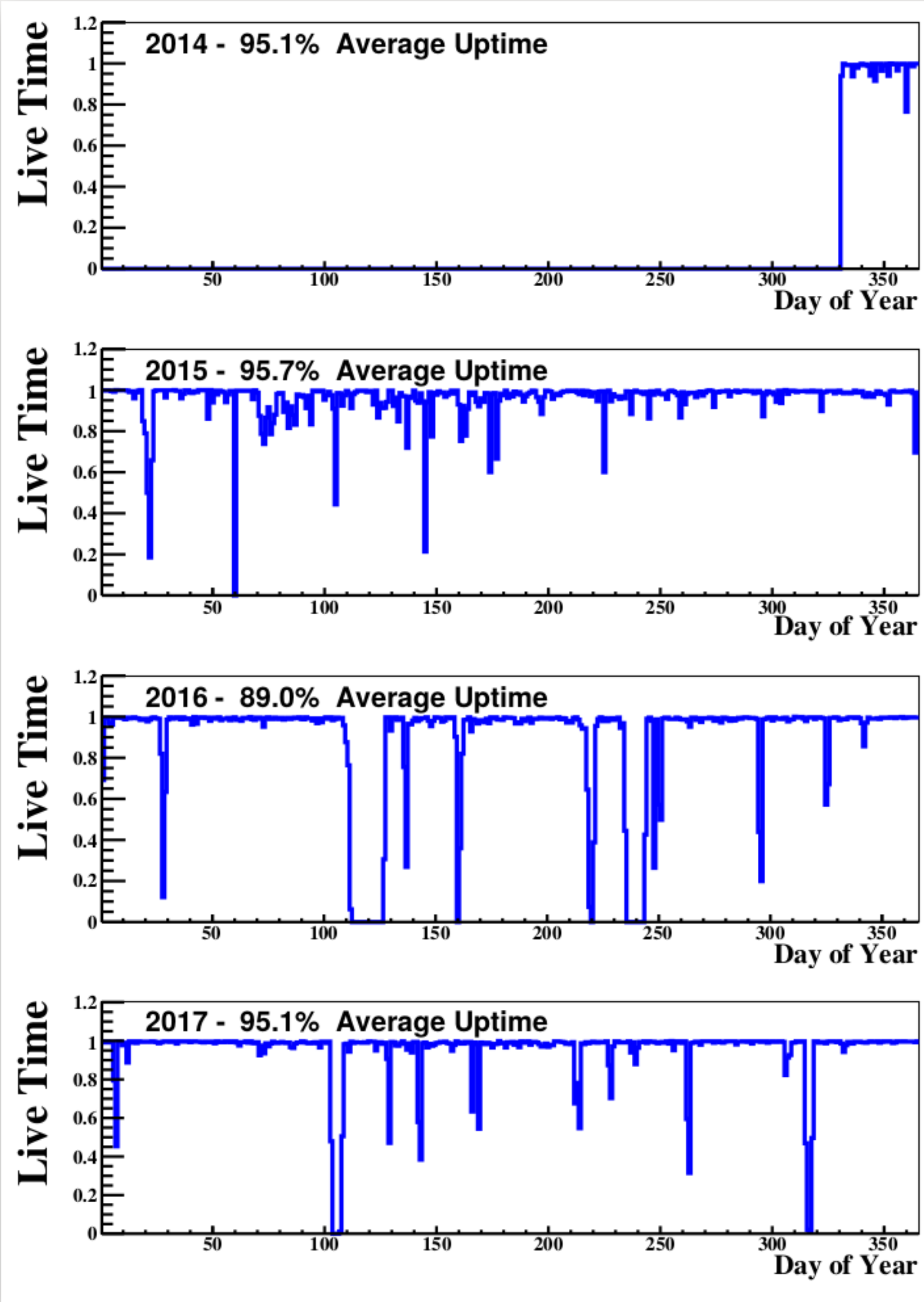}
\centering\includegraphics[trim={.1cm 0 0 0},clip, width=0.4\textwidth]{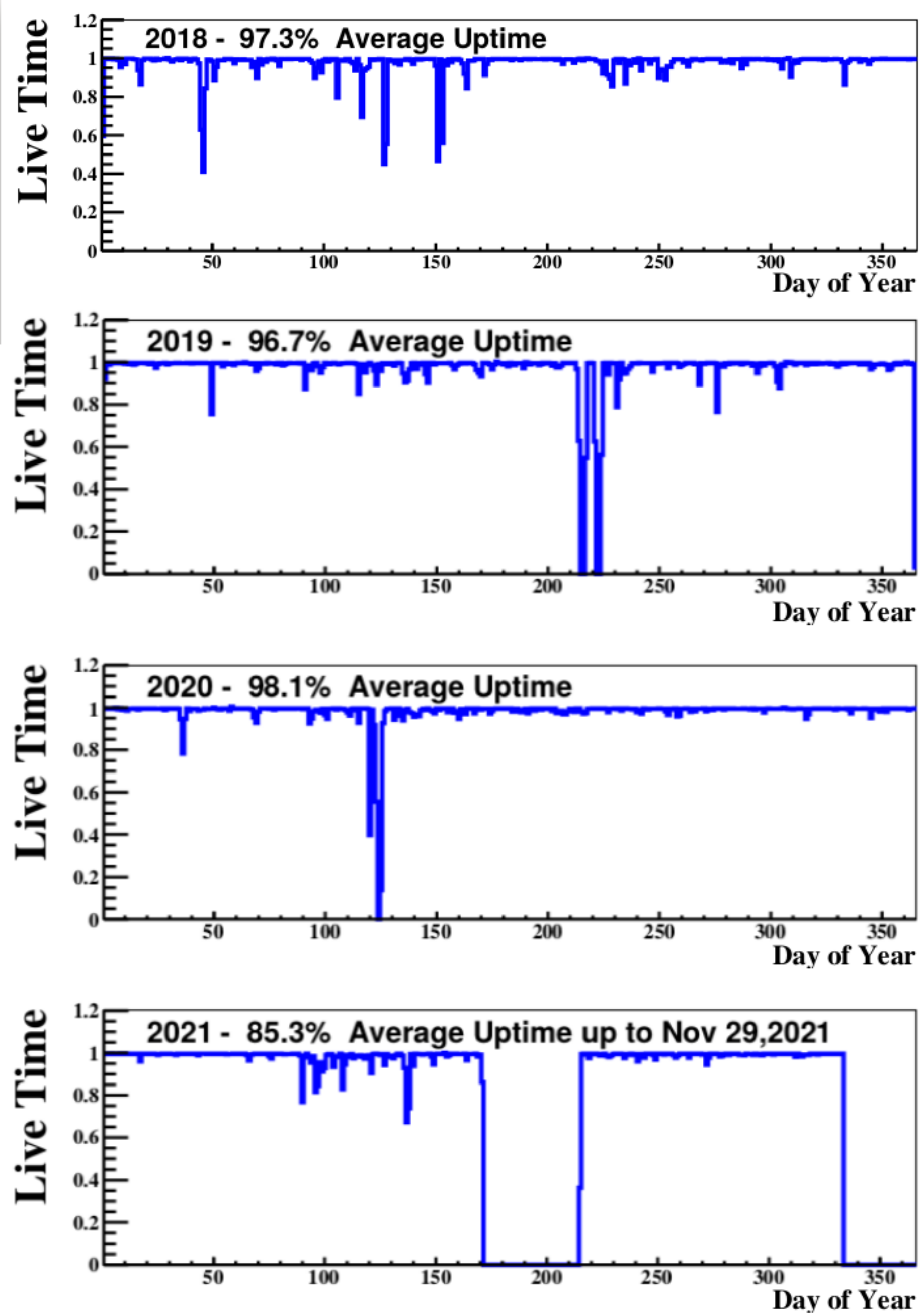}
\caption{Live time fraction plots of HAWC since the beginning of engineering operations with 250 WCDs.}
\label{fig:livetime}
\end{figure}

The live time fraction for each day since the beginning of engineering operations with 250 tanks on November 27, 2014 up to November 29, 2021 is shown in Figure~\ref{fig:livetime}. Over these 7 years, HAWC was operational and collecting data for more than 94\% of the time. The main source of downtime is power outages, mainly caused by lightning during the spring and summer seasons. HAWC was twice off for long periods: the first was in April of 2016 and the second in June and July of 2021. In both cases, lightning damaged critical transformers. We recently upgraded the electrical protection of the main transformer.

The COVID-19 pandemic tested the efficiency of remote HAWC operations. This pandemic forced us to work with minimal personnel. HAWC data-taking operation was not interrupted, and detector maintenance was adequately addressed during this unprecedented time. 

\begin{figure}[!htb]

\centering\includegraphics[width=0.5\textwidth]{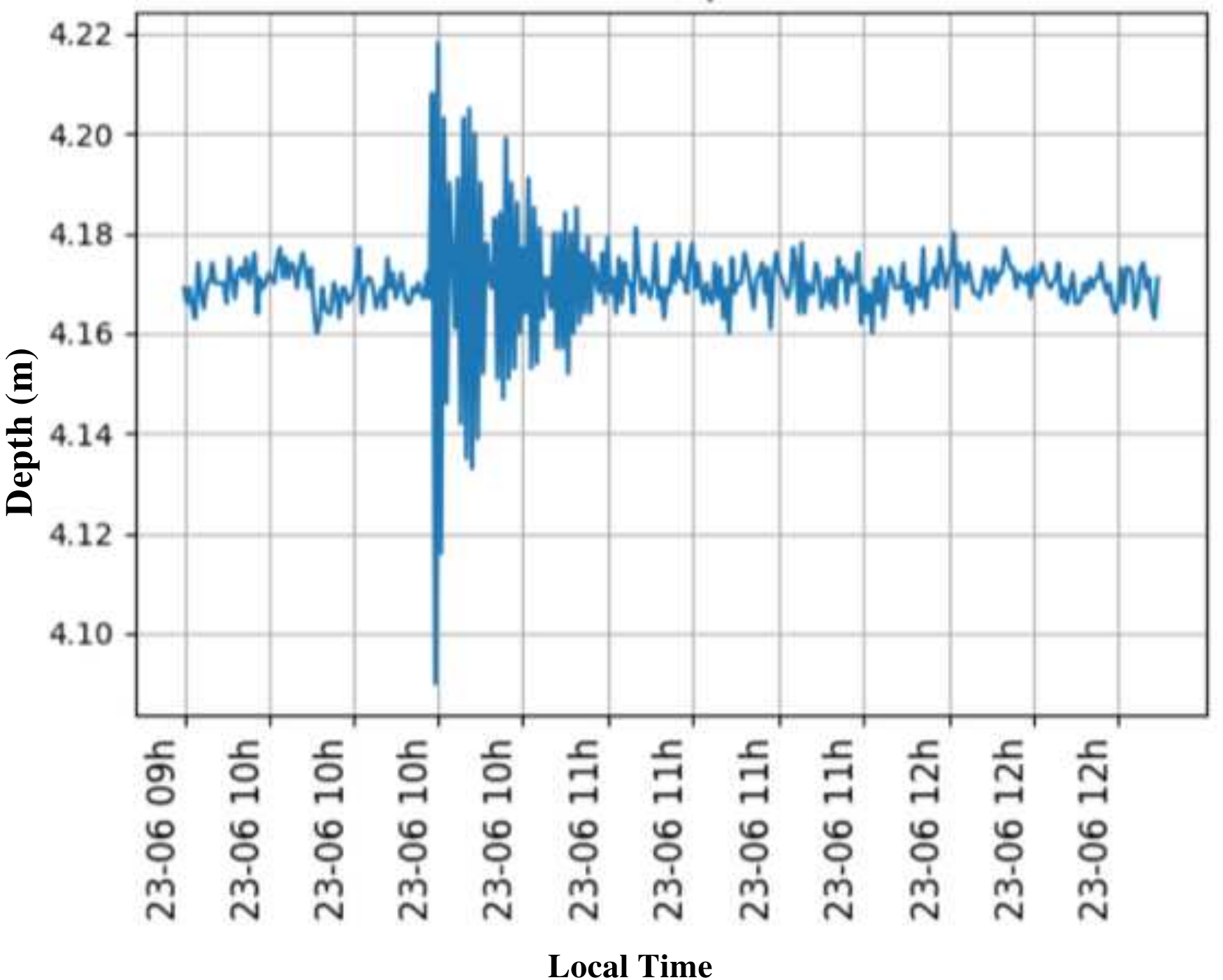}
\caption{Signal of the June 23, 2020 earthquake in M\'exico water level measured by the HAWC WCD E12.}
\label{fig:earthquake}
\end{figure}

HAWC also continued operations despite earthquakes which shook the site. On June 23, 2020, a 7.5 Richter magnitude earthquake hit M\'exico with an epicenter about 400 km South of Sierra Negra. Effects near the site were at level V on the Modified Mercalli scale. No damage in HAWC was observed, even though the WCDs registered a response to the earthquake in the scaler data and water level sensors. We present an example of the response of a water level sensor showing the sloshing of the water in a WCD in Figure \ref{fig:earthquake}. This event confirms the robustness of HAWC construction.

In January 2021, an overhead internet optical fiber was damaged between the site and the base camp at the foot of the mountain where our removable disk drives are located.  Local data caching in the CH allowed us to continue data taking for three weeks during the repair.  HAWC ran successfully without remote monitoring or intervention. We recently added a Starlink \cite{Starlink}) internet ground station to provide a reliable backup path for communications and monitoring.

The high altitude also challenges the health and cognitive abilities of humans, especially those coming to the site from low altitude. HAWC provided personnel training on the awareness of the symptoms of altitude sickness and implemented policies and procedures to help ensure safety.  Some of these procedures included making safety equipment such as a satellite phone, supplemental oxygen and pulse oximeters available at the site. Safety policies included the use of two-person rules, site visitation logs and required medical waiver forms that informed personnel of their risks.

\subsection{Networking and Site Computers}
\label{sec:networking}

The critical site network components run on UPS or solar power backup, particularly the main switches, router, hardware, the environmental monitoring system or EMS (Section \ref{S:ems}), and the site monitoring computers. The solar backup (Section 3.4) maintains basic connectivity during power outages and enables remote recovery after power is restored, as the EMS computer is employed to reboot other systems and computers.  For personnel security on site, our solar photovoltaic system powers the internet, satellite phone, and WiFi. Remote login to the site proceeds by VPN (Virtual Private Network).

A fiber directly connects the site internet to the base camp at the foot of the mountain. The connection from the base camp to the computer center at the Instituto de F\'isica of the Universidad Nacional Aut\'onoma de M\'exico (UNAM) initially included a microwave link. However, it was insufficiently reliable for remote monitoring and control.  We now have a commercial 40Mb/s fiber link with excellent interactive response. This comfortably handles monitoring data, and we have even updated control room operating systems with this connection, but it is not able to handle raw data transfers (See Section \ref{sec:data_transfer}).  

Standard computer disks do not operate reliably above $\sim$ 3 km (10,000 ft) because their design requires sufficient air density for cooling and to support the disk heads above the rotating platter. Standard disks are sufficient at the 2689 m altitude base camp (see Section \ref{sec:data_transfer}). The larger HAWC site data storage disks with heavy rewrite cycles are He-filled disks designed for computer-center use; they function reliably at HAWC altitude. We use fast SSDs to hold the operating system in all HAWC site computers, and for disks smaller than 1TB where cost is not prohibitive.  

The cooling issues for site computers are dealt with by maintaining low CH temperatures, preferring low-power computers, or auxiliary fans.  Site computers include low-power computers for the EMS system, monitoring, DAQ control, and calibration;  six computers carry out the reconstruction and a dedicated archive computer stores the raw data. A separate server runs real-time science analysis tasks using reconstructed data.

\subsection{The Environmental Monitoring System (EMS)}
\label{S:ems}

The EMS system performs low-level environmental monitoring; its components provide inputs to the monitoring system described in the next section. The EMS includes a weather station, an electric field monitor, temperature and voltage monitoring, water level monitoring for the WCDs, and webcams. The weather station is a Campbell Scientific Datalogger CR1000. We record atmospheric pressure, wind speed and direction, temperature, relative humidity, rain, and solar radiation every minute.  The logging software was developed by ADVANTECH \cite{ems}.

Strong electric fields, even without lightning, can cause elevated PMT count rates. We monitor the electric field with a Boltek EFM-100C RS485 Electric Field Mill. We installed it above the metal HAWC utility building, which is a good conductor and does not distort measurements. The high humidity and low temperatures at the site initially froze the moving parts of the field mill, so we added a warming circuit.  We read out the detector over optical fiber to avoid grounding issues.

The hardware for the temperature and voltage monitoring in the CH is based on two Advantech autonomous data acquisition boards, an ADAM-6015-7 channel RTD input module \cite{ADAM6015}, and Balco 1000 RTD thermistors as temperature sensors \cite{Balco1000}. These are placed on the top of each crate of high-power FEBs, for which it is essential to track temperature. We also monitor the temperature of the calibration and electronics rooms of the CH. An ADAM 6017 digitizes low voltages from the FEB power supplies.  Every 10 seconds, the 6017 also digitizes the external AC voltages through a rectifier circuit.  The EMS also records the state of the UPS  system and battery charge level. 

We monitor the water level in tanks, because some WCD bladders developed slow leaks. The water level monitoring system is based on an MPX4250AP pressure sensor \cite{MPX4250AP} placed at the bottom of the WCDs water volume. Each sensor requires three connections: a 5V power supply voltage along with ground and signal connections.  In order to prevent lightning-related transient over-voltages, protection diodes are required on the monitoring cables before entering the WCDs. An ethernet cable spark gap and another diode prevent over-voltages where the cables enter the CH. The signal ranges from 1 to 2 V, equivalent to a water depth between 0 to 5 m.  Signals, linearly proportional to absolute pressure, are multiplexed by a MEGA Arduino board to a Labjack U3-HV voltmeter. We record water levels with a 10-minute cadence for most tanks and every minute for a few tanks. Care was taken to use USB or RJ45 connectors in this system to avoid searching for exotic connectors at a remote site (see Section \ref{S:site} for more on the cabling).  While much less expensive than commercial sensors, up to 8\% of our water level sensors fail per year, usually due to compromised epoxy potting around the pressure sensor. As a final aspect of our monitoring, we periodically store images from webcams inside the CH and outdoors looking at the WCDs.

\subsection{Monitoring}
\label{monitoring}

Maintaining HAWC’s 95\% duty cycle requires a system specifically designed for remote monitoring. Monitoring data must be transferred off-site in an efficient manner that is robust to potential network outages. Scientists on the monitoring shifts must be able to view diagnostic information in a clear and user-friendly way that updates in real time so that experts can be notified of potential issues. Experts also require readily available tools to obtain more in-depth information and archival monitoring data to check abnormal behavior and validate candidate gamma-ray transients. Two software packages were developed to meet these goals. The Advanced Tracking of HAWC Experiment Notifications and Alerts (ATHENA) system handles data collection. The HAWC Observatory Monitoring for Experiment and Reconstruction (HOMER) system provides the monitoring user interface. 

ATHENA is a Python library which polls experimental hardware and stores the information in an SQL database. While the polling routine is different for each monitored device, ATHENA deals with all formatting in the back end, so data source experts can ignore database details. ATHENA automatically structures the data in a standardized way and has a built-in version control system to define a data stream change. This approach allows for new devices by simply adding a polling script that registers a new data stream with ATHENA. Most components are polled every minute, giving an effectively real-time view of the detector’s health. ZeroMQ \citep{zeromq} handles communication between processes reading data and inserting the data into the SQL database. We write approximately 8 GB of monitoring data per year. 

ATHENA moves monitoring data to the off-site computers. A custom Python database synchronization script compares on-site and off-site versions of tables, and 
the time column of each row. It then writes an SQL query to copy only new entries. Tables are synchronized one at a time. The synchronization launches at nearly the same cadence as the monitoring polling. Synchronizations time out and re-launch with approximately the same frequency (with a more permissive timeout for the larger tables), preventing a momentary bad connection or a slow table update from interrupting the entire monitoring process. After a prolonged network outage, we run scripts to move data in smaller packets to re-synchronize the monitoring more efficiently.

These optimizations mean approximately only 20 KB of data must be sent during each synchronization instance to keep everything up to date, which is well within the tolerance of the network. Unexpected outages can cause the sync routines to hang. A second master script running under Crontab automatically kills and restarts hanging sync scripts. This process has become less critical with the advent of a more reliable internet connection. 

HOMER is a collection of PHP web pages. It uses SQL wrapper libraries to sort the ATHENA database into arrays. These arrays are then either parsed for summary information or passed into a Google Charts API for plotting. While this API introduces an external dependency, it makes plots optimized for readability and provides interactivity to zoom in or out or to get the exact value of a data point by dragging the mouse over it. 

The top of the HOMER home page summarizes the most critical information and links to sub-pages with more detail. The critical information includes run status, temperatures of the electronics, scaler rates, and high voltage status. The home page is intuitive with color codes to indicate whether a component is in a normal state. HOMER also color-codes data if it is stale.

A specialized local version of HOMER exists at the site with reduced features. It is lighter-weight and independent of the Google APIs, allowing it to run even if the network connection is down. It is only used by experts at the site, and optimized for reliability and conciseness. The home page
contains only the critical information and links to on-site troubleshooting data such as scaler rates and electronics status. The site version of HOMER runs on the same machine as the on-site database so it reflects detector information in real-time up to the polling frequency. This system has proven vital during extended network outages. It also contains pages with JavaScript functions to turn high voltage channels on and off which directly and instantly access the high voltage modules to display status. These features are only available in the on-site version of HOMER to prevent HV channels being toggled by an off-site user or by a security breach at off-site computers.

An additional monitoring page outside the HOMER framework emphasizes data quality rather than hardware and environmental monitoring. The web page includes static figures and HTML text generated at the site and transferred to the off-site monitoring application using \textit{rsync}. The plots provide a light-weight snapshot of analysis and detector status updated every 20  minutes and include results from the online analysis, such as the current significance of bright gamma-ray sources. The page also provides more technical diagnostic plots of hardware based on higher-level reconstruction results. 

The final component of HAWC remote monitoring is the alert system integrated with the chat platform Slack\footnote{https://slack.com}. Python processes poll various DAQ components; abnormal responses send alerts to the HAWC Slack workspace. Alert scripts of this type monitor the high voltage, low voltage, run status, AC voltage, FEB temperatures, output file sizes, and data copying. Each alert sends pertinent details (such as which high voltage channels tripped) if a problem is detected. These messages alert experts to critical issues in real time and allow a quick response to correct problems.  To prevent overloading the Slack channels, alerts are temporarily silenced after a certain number of repetitions.

\subsection{Control of Cooling and Power}

The HVAC (Heating Ventilation and Air Conditioning) system has the primary responsibility for maintaining CH temperature.  The front-end boards are ECL based and require high power: the FEB crates together dissipate over 5kW, and the entire HAWC counting house consumes approximately 25kW of power.  Thus temperature control in the CH is critical.

Separate from the monitoring system, a  bimetallic safety interlock switch resides on top of each FEB crate; it turns off the DC power when the temperature exceeds $32^\circ$C. This can happen within seven minutes if we lose the AC power to the HVAC (Heating Ventilation and Air Conditioning) system while the crate is powered by UPS. Since overheating can occur quickly, the EMS monitoring of temperature and low voltage power is critical to provide information to the control scripts.  

HAWC is in a  challenging environment for air conditioning: the outdoors is often cooler than inside the CH, which stresses the air conditioning and could freeze the condenser.   The HVAC system removes most of the heat from the CH, but we supplement the HVAC with a water heat exchanger system described below.  This heat exchanger slows the rate at which the HVAC cycles on to half the time, thus extending the HVAC working life.  The heat exchanger system also allows the analog part of the front-end boards to operate more stably in a narrower temperature range.

A dual-loop heat exchange system performs supplemental heat removal. The heat exchanger is a Coolflow System III from Neslab with a maximum cooling capacity of up to 70 kW.  One of the working WCDs near the CH provides the large heat sink of $3^{\circ}$C water.  This cool water  circulates through one side of the heat exchanger, returning to the WCD slightly warmer water.  The secondary water loop is maintained at 6 to 12 $^{\circ}$C by controlling the flow rate to ``radiator'' units in each of the four high-power electronics racks, where air warmed by the electronics exchanges heat with the water from the secondary loop.  Each of the 4 radiators removes 6kW of heat.

Remote shifters can determine when the cooling system has tripped by viewing a webcam aimed at a LED indicator of the heat exchanger system. However, resetting after a trip is an expert task: it is necessary to first verify that the trip was not caused by a water leak, by checking the water sensors under each heat exchanger and under the pump.

\subsection{Slow Control}
\label{slowcontrols}

Slow control refers to operations related to control of the electronics which take place on the time scales (seconds or longer) much longer than the real-time DAQ scale needed for handling data from individual events.

\subsubsection{Automatic Run Restart}

HAWC runs normally last for 24 hours, after which the run is restarted automatically. However, the DAQ can crash when high electric fields near the site cause elevated PMT rates.   Automatic scripts detect crashes (whatever the cause), end a crashed run, reset the system, and restart a run without remote interaction, continuing this cycle until conditions are stable enough that attempted runs no longer crash.  This minimizes downtime during stormy weather.

\subsubsection{Power Outage Bridging and Shutdown}

HAWC uses two 30kVA (UPS) to bridge short power interruptions and allow automatic controlled power-down scripts to act for more prolonged interruptions, or in case of climate control failures leading to temperatures outside of the safe range in electronic crates.  Under regular operation, the UPS conditions AC input power, thus supplying ``clean'' power to the electronics; unconditioned (``dirty'') AC power is restricted to less-critical uses such as running HVAC, fans, soldering stations, lighting, and powering non-DAQ computers for on-site users outside  the electronics room.

When the external AC power voltage is less than 100 V, and after HAWC has been powered by UPS power for 2 minutes, scripts detect the loss of AC power. These scripts stop the data-taking run, turn off DAQ power supplies and power distribution units (PDUs) controlling non-internet-controllable items such as cooling fans, then turn off all DAQ computers in a controlled fashion. The scripts leave the monitoring system, internet switches, and the monitoring computer running under the power provided by the solar-power system and its separate UPS.  Since the main UPS could run the system for 40 minutes, it has more than sufficient power storage to safely restart the site as needed.

\subsubsection{Power Outage Recovery}

The Slack Alert system ensures that experts know when we lose AC power. Experts use the remote monitoring to detect when AC power has returned. They  run restart scripts to turn HAWC back on after power or HVAC is restored. The SNMP protocol restarts DAQ computers,  turns on cooling fans via remote-controlled PDUs and then  the DAQ low and high voltage power supplies. Developing and testing the shutdown and restart scripts took over a year to achieve smooth operation, but it has been critical to maintaining HAWC data taking with minimal downtime.

\subsection{Data Transfer}
\label{sec:data_transfer}

HAWC records data at a rate of approximately 2 TB/day (20 MB/s), or 730 TB per year. The site's primary disk array storage capacity is 82 TB, which provides storage for about 40 days of data recording. The data transfer from the HAWC observatory to the closest data center located at UNAM is carried out by transporting portable disk arrays (PDAs). Each disk array has a capacity of 29TB or 36TB. HAWC has 12 of these PDAs. We transfer data down the mountain on a 15km 1 Gb optical fiber to the nearby town of Atzitzintla, where our base camp is: an office of the Instituto Nacional de Astrof\'isica, \'Optica, and Electr\'onica (INAOE). This office is set up to handle the data transfer to the PDAs and make backup copies of the data. 

We can store up to 110 days of data and backup copies in the office or at the site.  We ship the PDAs to UNAM approximately every six weeks.
The local storage ability made it possible for the HAWC experiment to continue operating during the COVID-19 pandemic despite its effects on the regular schedule for data movement. 

Once the PDAs arrive at UNAM, they are backed up and checked against on-site data. This first identical copy stays in permanent storage at the UNAM data center. Afterwards, the data  moves via a 10 Gb/s connection (of which HAWC uses less than 10\%) to a second data center located at the University of Maryland. Once the data is checked and archived at both data centers, it is removed from the site. In case of data loss at one facility, data can be recovered from the copy at the second location.  

\subsection{Real-time Science Programs}
\label{sec:real_time}

HAWC carries out several science analyses in real time at the HAWC site, based on the online reconstruction. We search for transients of the Crab, Mrk 421 and Mrk501 daily, search the sky for bright sources on time scales from hours to 3 days, and search for GRBs (independently, or based on satellite alerts) on time scales between 0.3 and 1000 seconds. Automatic email alerts are sent to partners that have an agreement with HAWC. We post Astronomer's Telegrams (ATels) and Gamma-ray Coordinates Network (GCN) circulars for interesting alerts. We send events and alerts to Astrophysical Multimessenger Observatory Network (AMON)~\citep{hugo19} from our independent GRB search. Anything below a false-alarm rate of 12 per year is sent as a Notice to GCN. The latency is less than a minute. 

\section{The Upgrade of HAWC} 
\label{S:Future}

While the gamma/hadron discrimination method in HAWC works well, it performs better with increasing gamma-ray energy, with the best performance at the highest energies. Additionally, a significant fraction of events passing the multiplicity trigger criterion have their shower core outside the 300-WCD array.  While gamma/hadron discrimination is possible for these events, without good knowledge of the position of the shower core, there is ambiguity in the reconstruction of the shower direction and shower size (needed to determine the energy of the primary gamma ray or cosmic ray).

HAWC, like Milagro before it, has been upgraded by adding outrigger detectors. This expanded array, installed between 2016 and 2018, enhances the sensitivity above 10 TeV by increasing the area over which shower cores can be detected by a factor of 3-4 and improving the angular resolution (see \cite{Marandon2021} and references therein). The details of this outrigger array will be published in a forthcoming paper.

\section*{Acknowledgement}

We acknowledge the support from: the US National Science Foundation (NSF); the US Department of Energy Office of High-Energy Physics; the Laboratory Directed Research and Development (LDRD) program of Los Alamos National Laboratory; Consejo Nacional de Ciencia y Tecnolog\'ia (CONACyT), M\'exico, grants 271051, 232656, 260378, 179588, 254964, 258865, 243290, 132197, A1-S-46288, A1-S-22784, c\'atedras 873, 1563, 341, 323, Red HAWC, M\'exico; DGAPA-UNAM grants IG101320, IN111716-3, IN111419, IA102019, IN110621, IN110521, IN102223 ; VIEP-BUAP; PIFI 2012, 2013, PROFOCIE 2014, 2015; the University of Wisconsin Alumni Research Foundation; the Institute of Geophysics, Planetary Physics, and Signatures at Los Alamos National Laboratory; Polish Science Centre grant, DEC-2017/27/B/ST9/02272; Coordinaci\'on de la Investigaci\'on Cient\'ifica de la Universidad Michoacana; Royal Society - Newton Advanced Fellowship 180385; Generalitat Valenciana, grant CIDEGENT/2018/034; Chulalongkorn University’s CUniverse (CUAASC) grant; Coordinación General Académica y de Innovación (CGAI-UdeG); Cuerpo acad\'emico PRODEP-SEP UDG-CA-499. H.F. acknowledges support by NASA under award number 80GSFC21M0002. We also acknowledge the significant contributions of former members of the HAWC collaboration: R. Arceo, E. Almaraz, M. Alvarez, J. R. Angeles Camacho, B. M. Baughman, D. Berley, P. Colin Farias, M. A. Diaz Cruz, D. W. Fiorino, D. Garcia, V. Grabski, Z. Hampel-Arias, J. Jablonski, N. Kelley-Hoskins, M. Lamprea, S. S. Marinelli, J. Martínez, C. Rivière, M. Rosenberg, P. Vanegas, X. J. Vázquez, O. Vázquez-Estrada, G. Vianello, D. Warner, T. Weisgarber,  T. Yapici, and P. W. Younk. Thanks to Scott Delay, Luciano D\'iaz, Eduardo Murrieta and Janina Nava for technical support.




\end{document}